\documentclass{aa}  
\usepackage{color}
\usepackage{ctable}
\usepackage{tablefootnote}
\usepackage{natbib,twoopt}
\usepackage{natbib}
\bibpunct{(}{)}{;}{a}{}{,} 
\usepackage{graphicx}
\usepackage{amsmath,amssymb}
\usepackage{threeparttable}

\usepackage{titlesec}
\bibpunct{(}{)}{;}{a}{}{,} 
\makeatletter
\newcommandtwoopt{\citeads}[3][][]{\href{http://adsabs.harvard.edu/abs/#3}%
{\def\hyper@linkstart##1##2{}%
\let\hyper@linkend\@empty\citealp[#1][#2]{#3}}}
\newcommandtwoopt{\citepads}[3][][]{\href{http://adsabs.harvard.edu/abs/#3}%
{\def\hyper@linkstart##1##2{}%
\let\hyper@linkend\@empty\citep[#1][#2]{#3}}}
\newcommandtwoopt{\citetads}[3][][]{\href{http://adsabs.harvard.edu/abs/#3}%
{\def\hyper@linkstart##1##2{}%
\let\hyper@linkend\@empty\citet[#1][#2]{#3}}}
\newcommandtwoopt{\citeyearads}[3][][]%
{\href{http://adsabs.harvard.edu/abs/#3}
{\def\hyper@linkstart##1##2{}%
\let\hyper@linkend\@empty\citeyear[#1][#2]{#3}}}

\def\kms{$\mathrm{km\,s}^{-1}$}

\def\kms{$\mathrm{km\,s}^{-1}$}

\def\Msun{\ensuremath{\,M_\odot\,}}
\def\Lsun{\ensuremath{\,L_\odot\,}}
\def\Rsun{\ensuremath{\,R_\odot\,}}

\def\Mdot{\ensuremath{\dot{M}\,}}

\makeatother

\usepackage{txfonts}
\begin{document}

\title{Wind-envelope interaction as the origin of the slow cyclic brightness variations of luminous blue variables}
   \author{L. Grassitelli\inst{1}\fnmsep\thanks{e-mail: luca@astro.uni-bonn.de},        N. Langer\inst{1,2}, J. Mackey\inst{3,4}, G. Gr\"afener\inst{1}, N.J. Grin\inst{1}, A.A.C. Sander\inst{5}, \& J.S. Vink\inst{5}}

\institute{Argelander-Institut f\"ur Astronomie, Universit\"at Bonn, Auf dem H\"ugel 71, 53121 Bonn, Germany
\and
 Max-Planck-Institut f\"ur Radioastronomie      , Auf dem H\"ugel 69, 53121 Bonn, Germany
\and
Dublin Institute for Advanced Studies, 31 Fitzwilliam Place, Dublin 2, Ireland
\and
Centre for AstroParticle Physics and Astrophysics, DIAS Dunsink Observatory, Dunsink Lane, Dublin 15, Ireland
\and
 Armagh Observatory and Planetarium, BT61 9DG Armagh, College Hill, Northern Ireland
}
\abstract{
Luminous blue variables (LBVs) are hot, very luminous massive stars displaying large quasi-periodic variations in brightness, radius, and photospheric temperature on timescales of years to decades. The physical origin of this variability, called S\,Doradus cycle after its prototype, has remained elusive. We study the feedback of stellar wind mass-loss on the envelope structure in stars near the Eddington limit. We calculated a time-dependent hydrodynamic stellar evolution, applying a stellar wind mass-loss prescription with a temperature dependence inspired by the predicted systematic increase in mass-loss rates below 25\,kK. We find that when the wind mass-loss rate crosses a well-defined threshold, a discontinuous change in the wind base conditions leads to a restructuring of the stellar envelope. The induced drastic radius and temperature changes, which occur on the thermal timescale of the inflated envelope, in turn impose mass-loss variations that reverse the initial changes, leading to a cycle that lacks a stationary equilibrium configuration. Our proof-of-concept model broadly reproduces the typical observational phenomenology of the S\,Doradus variability. 
 We identify three key physical ingredients that are required to trigger the instability: inflated envelopes in close proximity to the Eddington limit, a temperature range where decreasing opacities do not lead to an accelerating outflow, and a mass-loss rate that increases with decreasing temperature, crossing a critical threshold value within this temperature range. Our scenario and model provide testable predictions, and open the door for a consistent theoretical treatment of the LBV phase in stellar evolution, with consequences for their further evolution as single stars or in binary systems.

}

\authorrunning{Grassitelli et al. 2020}
\titlerunning{Physical origin of the S Doradus cycle in luminous blue variables}
\maketitle       
\section{Introduction}

Luminous blue variables (LBVs) are a class of massive stars showing dramatic spectroscopic and photometric variations, as well as strong and variable stellar wind mass-loss \citep{1988Lamers,1989Lamers,1994Humphreys,2001vanGenderen,2012Vink}. 
This class encompasses stars displaying variability on different timescales and of drastically different intensity \citep{1996deKoter,2018Kalari}. Microvariability in LBVs refers to $\lesssim 0.1\,$mag variations on a timescale of days to weeks, consistent with the dynamical timescale of B-supergiant stars, and likely related to heat- or convection-driven nonradial waves \citep{2007Lefever,2018Jiang} that are excited in the stellar envelope. 
The S Doradus (S Dor) variables are the typical LBVs, characterized by $\approx 0.5-2\,$mag quasi-periodic variations on a timescale of years to decades, consistent with the thermal timescale of blue supergiant envelopes \citep{1992Maeder}. Approximately one hundred massive stars are known to undergo such cyclic brightness variations, from environments with sub- and suprasolar metallicity \citep[e.g.,][]{1995Lamers,1997Crowther,2000Massey,2013Humphreys,2015Sholukhova}. 
Giant eruptions instead are $\gtrsim 1-2\,$mag brightness variations associated with an episode of high mass-loss \citep[][]{1997Davidson,2011Smith}. Only two stars in the Milky Way have been reported to have potentially undergone episodic mass-loss like this \citep[$\eta$-Carinae and P-Cygni, which ejected $\approx 10\Msun$ and $\approx 0.1\Msun$, respectively, with a few more candidates in other galaxies,][]{2011Smith,2020Moriya}.
Because it is so rare, no direct evidence suggests that this phenomen is a common phase in the evolution of massive stars.  

The different phenomenology of these variations suggests a diverse origin; we focus on the typical LBV S Dor variables here. 
Several hypotheses have been made to interpret the physical origin of the S Dor phenomenon, invoking turbulent pressure and subsurface convection, pulsations, instability of density inversions, binarity, and the generic proximity to the Eddington limit \citep[for a review, see][]{1994Humphreys,2012Vink}.
No theoretical model so far provides a consistent mechanism that can reproduce the origin of their characteristic quasi-periodic variations in radius (hundreds of solar radii) and temperature (from $\rm\approx 30$ to 10\,kK). 

This phase is likely crucial in the poorly established evolution of massive stars. A significant fraction of the envelope is potentially lost during this phase. The lack of very luminous red or yellow supergiants \citep{1994Humphreys} even at low metallicities indicates that the S Dor phase might fundamentally contribute to deplete the H-rich outer layers of massive stars, leading to a blueward evolution toward the Wolf-Rayet (WR) regime \citep{1994Langer,1997Maeder}. This would inevitably affect predictions concerning the main- and especially the post main-sequence evolution of single stars and binary systems at different metallicities. 
Implications involve an increased kinetic feedback onto the interstellar medium, providing early feedback that affects star formation and galaxy morphology \citep{2013Stinson,2014Hopkins}, a reduction in the angular momentum budget of rotating massive stars, affecting statistics of long gamma-ray bursts \citep{2006Yoon,2020Chrimes} and the final black hole masses, the position in the Hertzsprung-Russell (HR) diagram, and the upper luminosity limit of red and yellow supergiants \citep{1995Nieu,1998deJager,2018Davies,2020Higgins}, as well as the rate and ratio of supernova subtypes \citep{2003Heger} and the number of WR stars \citep{2012Langer}. Both the extreme phenomenology and the still unaccounted-for effect that the S Dor phase has on evolutionary calculations make of this one of the most puzzling stages in the evolution of massive stars.
In this work, we employ newly developed 1D hydrodynamically self-consistent stellar models with boundary conditions set at the sonic point to investigate the interplay between low-density envelopes, accelerating outflows, and variable mass-loss rates by stellar winds in models near the Eddington limit. 

\section{Methods}\label{method}

We adopted the Bonn evolutionary code \citep{2000Heger,2006Yoon,2011Brott}, a 1D Lagrangian hydrodynamic stellar evolutionary code that solves the hyperbolic set of partial differential equations that describe the stellar structure, with well-defined boundary conditions. These equations, together with the network of nuclear reaction rates, the set of equations of the mixing length theory for convection \citep{1958Vitense}, and the OPAL opacity tables \citep{1996Iglesias}, define the structure and evolution of a stellar model. In addition, mass-loss by stellar wind can also be applied \citep{2001Vink}, using a pseudo-Lagrangian scheme for the outer 50$\%$ of the stellar model \citep{1977Neo,2018Grassitelli}. 

First, we adopted the classical plane-parallel gray atmosphere boundary conditions and evolved a nonrotating 100$\Msun$ stellar model with composition [X,Y,Z]=[0.7,0.28,0.02], where X and Y are the hydrogen and helium mass-fraction, and Z is the metallicity, including mass-loss by stellar wind \citep{2011Brott}, until the model had Y=0.86 at the center and Y=0.48 at the surface. The model had a luminosity of $\rm log(L/\Lsun)\approx 6.25$, a radius of 300 $\Rsun$, a mass of 73\Msun, a mass-loss rate $\rm log(\Mdot/\Msun\,yr^{-1})\approx-3.1$, and an effective temperature of 10\,kK. We adopted a mixing length parameter of 2.5.

We then replaced the classical outer boundary conditions and applied sonic-point boundary conditions \citep{2018Grassitelli}, 
\begin{equation}\label{boundary}
\upsilon^2 = \frac{k_B T_s}{\mu m_H} = c_s^2\quad ,\quad
\Mdot=4 \pi r^2 \rho \upsilon \quad,
\end{equation}
where $\upsilon$ is the velocity in the radial direction, $r$ is the radial coordinate, $k_B$ is the Boltzmann constant, $\mu $ is the mean molecular weight, $m_H$ is the proton mass, $\Mdot$ is the mass-loss rate, $c_s$ is the isothermal sound speed, $T_s$ is the sonic point temperature, and $\rho$ is the density. We also implemented customized Rosseland opacities for the outer stellar layers, which modify the OPAL tables to include turbulent line-broadening (Sect.\ref{opacita}). The mass-loss rate can either be set manually, or with a mass-loss prescription.

In our hydrodynamic calculations, the given mass-loss rate imposed an outward-directed mass-flux throughout the stellar model, affecting the local force and energy balance while fulfilling conservation of mass \citep{2000Heger,2006Petrovic,2018Grassitelli}. In this respect, the stellar structure equations include and are affected by the inertial acceleration term (Eq.\ref{momentumeq}) and the local kinetic and advected energy of the outflow.
Each computed stellar model self-consistently adjusted its structure according to the imposed mass-loss rate until the resulting envelope structure was such that the mass-outflow reached the sonic point.

The considered model is thought to be representative of the typical parameters of an LBV, being a massive stellar model in a late evolutionary phase located in the top right corner of the HR diagram, with a massive envelope and helium enrichment at the surface. 
In Sect.\ref{Sect.steadymodels} and Sect.\ref{Sect.windenvelope} we introduce the structural effects that different mass-loss rates impose on such massive stellar models. 
In Sect.\ref{Sect.steadymodels} we compute a set of steady-state hydrodynamic stellar structure models in thermal equilibrium, with sonic point boundary conditions rather than the plane parallel gray atmosphere conditions. We start from the evolved stellar model with the classical outer boundary conditions we described above, and by using the sonic-point boundary conditions, compute stellar models in which different constant mass-loss rates are applied.

In Sect.\ref{Sect.windenvelope} we perform a time-dependent hydrodynamic stellar evolution calculation with sonic-point boundary conditions of the same evolved stellar model as above, covering $\approx 10^3\,$yr, but imposing a physically motivated temperature-dependent mass-loss prescription using time steps of $10^6$ seconds. We chose this time step because it is longer than the dynamical timescale of the model ($\approx 5\cdot 10^5$s) and shorter than the thermal timescale. For simplicity, we neglected rotation, turbulent pressure, and magnetic fields in our models. 
This time-dependent calculation is meant to be a proof-of-concept of a novel physical instability, based on a qualitative picture of the physical conditions encountered in the outer layers of LBVs (Sect.\ref{Sect.windenvelope}). We therefore consider the evolutionary history and the specific characteristic of the adopted stellar model of only secondary relevance. 
The intent is to investigate the appearance of instability, and later in the text, we highlight what we suggest are the necessary physical conditions to interpret our numerical results, and attempt an educated comparison to observations.

In the time-dependent calculation, we also adopt simplified wind models \citep{2018Grassitelli} to estimate the location of the photosphere and the optical depth of the sonic point. These wind models adopt a beta-velocity law with exponent unity \citep{2008Grafener}, terminal wind velocities proportional to the escape-speed times a factor 2.6 on the hot side and 1.3 on the cool side of the bistability temperature  \citep{1995Lamers}, and a temperature stratification given by a T-$\tau$ relation \citep{2002Nugis}. We implicitly assume in this manuscript that once the outflow becomes supersonic, the flow velocities do not become subsonic again at larger distances.

\section{Steady-state hydrodynamic massive star models}\label{Sect.steadymodels}

The extreme luminosities and mass-loss rates of LBVs unambiguously indicate that these stars are close to their Eddington limit, that is, the limit on the hydrostatic stability of stars given by the balance between the inward-directed gravitational force and the outward-directed radiative force \citep{1988Lamers,1992Maeder,1994Humphreys,2012Langer,2014Smith}.
Massive stars in proximity to the Eddington limit develop {\it \textup{inflated}} envelopes, that is, quasi-hydrostatic, radiation-pressure-supported envelope structures characterized by low densities, low heat capacities, and turbulent convective motion over a large radial extent \citep{1999Ishii,2006Petrovic,2012Grafener,2015Sanyal,2015GrassitelliA,2015Jiang,2015Owocki}. 
Massive stars also develop strong, radiation-driven winds that ar eaccelerated by momentum transfer from the intense radiation field to the atmospheric layers by scattering and line absorption of photons \citep{1999LamersCassinelli}.

Both inflated envelopes and radiation-driven winds appear in relation to outward-increasing opacities
in the outer stellar layers \citep{2002Nugis,2015Sanyal,2015Owocki,2018Grassitelli,2020Sander}.
However, the opacity of stellar matter in the outer layers is neither constant nor does it monotonically increase with radial coordinate. Rather, it shows pronounced local maxima, called opacity {\it \textup{bumps}}, associated with the recombination temperatures most notably of hydrogen at 10\,kK, helium-{\sc i} at 15\,kK, helium-{\sc ii} at 30\,kK, and iron at 150\,kK \citep[see Fig.\ref{opacita};][with the H- and He{\sc i}-bumps often appearing blended, thus indicated as H/He{\sc i}-bump]{1996Iglesias}.
Consequently, the radiative acceleration and the local proximity to the Eddington limit (Eq.\ref{gamma}) in the outer stellar layers are a nonlinear function of temperature and thus radial coordinate when the Rosseland opacity is considered (see also Appendix \ref{sect.opacita}).

We compute a set of steady-state hydrodynamic stellar structure models with sonic-point boundary conditions and adopt different constant mass-loss rates (Sect.\ref{method}). We study the readjustment of the stellar models to the adopted mass-loss rates, investigating with particular attention the physical conditions at which our hydrodynamic models find the early acceleration to reach the sonic point for the assumed mass-loss rate.

\subsection{Effects of mass loss on the envelope structure}\label{Sect.appendixmassloss}

\begin{figure}
\resizebox{0.9\hsize}{!}{\includegraphics{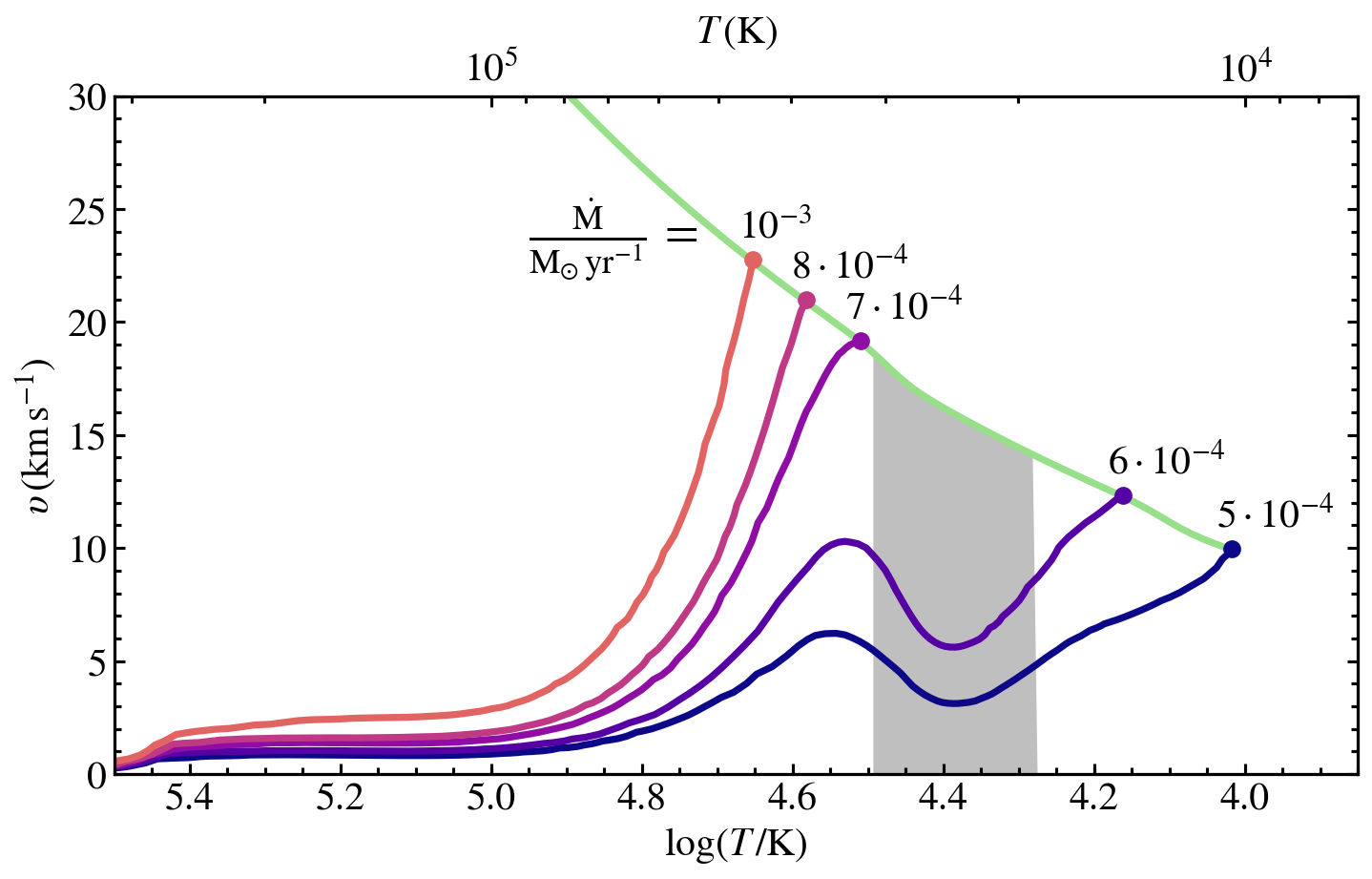}}
\resizebox{0.9\hsize}{!}{\includegraphics{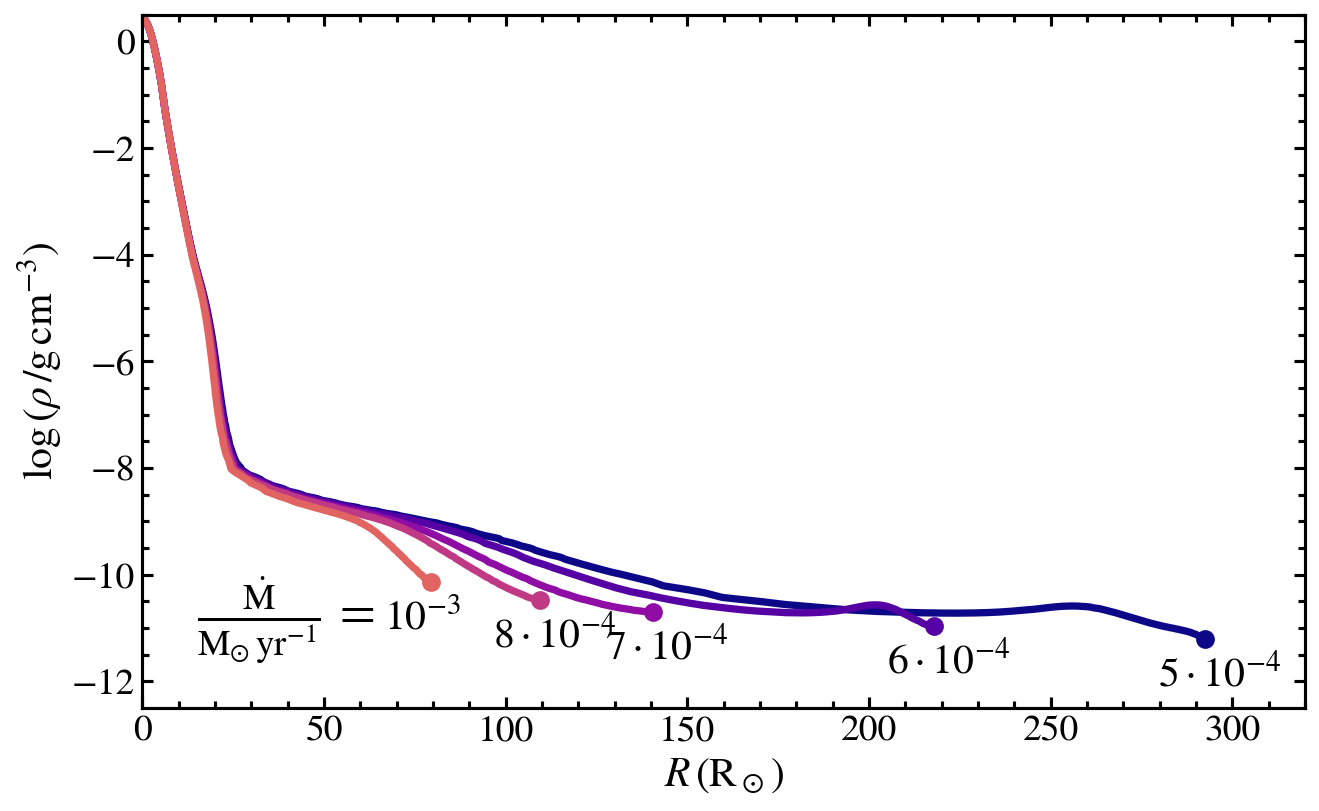}}
\caption{Velocity (upper panel) and density (lower panel) profiles as a function of temperature and radial coordinate, respectively, for a set of 73\Msun steady-state hydrodynamic stellar models with the boundary conditions set at the sonic point and with different applied mass-loss rates (in units of \Msun/yr, indicated by the numbers in the panels). The green line indicates the local sound speed, and the gray shaded area shows the forbidden temperature range. }
\label{structure}
\end{figure}

Figure \ref{structure} shows the velocity and density structure in the outer subsonic stellar layers of a set of the massive steady-state hydrodynamic stellar models in their late core-hydrogen-burning phase described in Sect.\ref{method}, where different constant mass-loss rates by stellar wind have been adopted. 
Our stellar models were computed up to the sonic point, assuming that once supersonic, the stellar wind does not decelerate. In the limit of validity of the diffusive approximation for radiative energy transport at the sonic point, the stationary subsonic stellar structure is independent of the detailed conditions in the supersonic wind \citep{2018Grassitelli}.

 In the investigated range of mass-loss rates, all the computed models present inflated envelopes supported by the radiation pressure gradient, and have a local Eddington factor $\Gamma \approx$ 1 at the sonic point (Eqs.\ref{gamma} and \ref{kappaedd}). In the context of stellar wind models and theory, $\Gamma \approx$ 1 at the sonic point suggests that these transonic outflows correspond to the base of winds driven by radiation pressure \citep{2007Shore}.
We can distinguish two families of solutions for the outer stellar structure: the more radially extended stellar models, with nonmonotonic velocity profiles that reach transonic velocity at $\rm \approx 10-15\,kK$, and the more compact stellar models with steeper monotonic velocity profiles and sonic-point temperatures in the range $\rm \approx 30-40\,kK$.

The first family of solutions, corresponding to stellar models with lower mass-loss rates (i.e., $\rm \Mdot< 7\times 10^{-4}\,$\Msun/yr), show inflated density stratification with low densities for extended regions of space, similar to those of purely hydrostatic stellar models with gray atmosphere boundary conditions. 
Our massive star models show a velocity increase in the proximity of the opacity bumps (Fig.\ref{opacita}), displaying at first a relative small increase in flow velocity around 200\,kK (i.e., around the Fe-opacity bump), followed by a more pronounced acceleration around 40\,kK. However, the acceleration at the temperature of the He{\sc ii}-bump is insufficient to reach transonic velocities in these more extended stellar models. Starting from $\approx 30$\,kK, the opacity decreases, and so do the flow velocities in Fig.\ref{structure}. Around $\approx 20$\,kK, the opacity increases once more, this time due to the recombination of H and He{\sc i}, which leads to transonic outflows with sonic-point temperatures of $\approx 10$\,kK.  
Stellar models experiencing higher mass-loss rates instead have higher outflow velocities, reaching the sonic point at temperatures $\rm T_s \approx 30-40$\,kK (Fig.\ref{structure}).
The higher mass-loss rates and the demand to conserve mass imply higherer mass-fluxes through the stellar envelopes, and therefore greater inertia. Our stellar models readjust accordingly, displaying steeper velocity gradients without encountering the sharp velocity decrease seen for lower mass-loss rates. For the higher mass-loss rates ($\rm \Mdot\geq 7\times 10^{-4}\,$\Msun/yr), our stellar models therefore locate the sonic point at smaller sonic-point radii and higher temperatures, following the increase in opacity associated with the He{\sc ii}-recombination.

The lower panel of Fig.\,\ref{structure} depicts the density profiles of our stellar models, illustrating the effect of different mass-loss rates on the derived outer structures. Models with a lower mass-loss rate have the typical structure of an inflated star, with a very extended low-density envelope at the Eddington limit \citep{2015Sanyal}. The resulting inflated envelope structure manifests as an almost flat density and temperature stratification for hundreds of solar radii, dramatically increasing the photospheric or sonic-point radii of massive stars near the Eddington limit.
In Fig.\ref{structure}, at the base of the envelope ($\approx 25\Rsun$), the radiative force from the Fe-bump is responsible for the initial inflation of the stellar model. Further inflation is then caused by the He{\sc ii}-bump (at $\approx 150\Rsun$), until a sonic point is reached, concomitant with the increase in opacity at $\approx 10$\,kK. Moreover, these models develop a slightly overdense region near the surface, that is, a density inversion, associated with the decrease in flow velocity \citep{1992Maeder,2012Grafener,2015Sanyal,2015Owocki,2018Grassitelli}.
On the other hand, models with a higher mass-loss rate require higher sonic-point densities, and thus are still initially inflated because of the Fe-bump, but find their sonic point at much smaller radii, without further envelope inflation due to the He{\sc ii}-bump. Their envelope structures shows steeper density and gas pressure gradients than the outer layers of the stellar models with lower mass-loss rates.

\subsection{Sonic-point conditions}
\begin{figure}
\resizebox{\hsize}{!}{\includegraphics{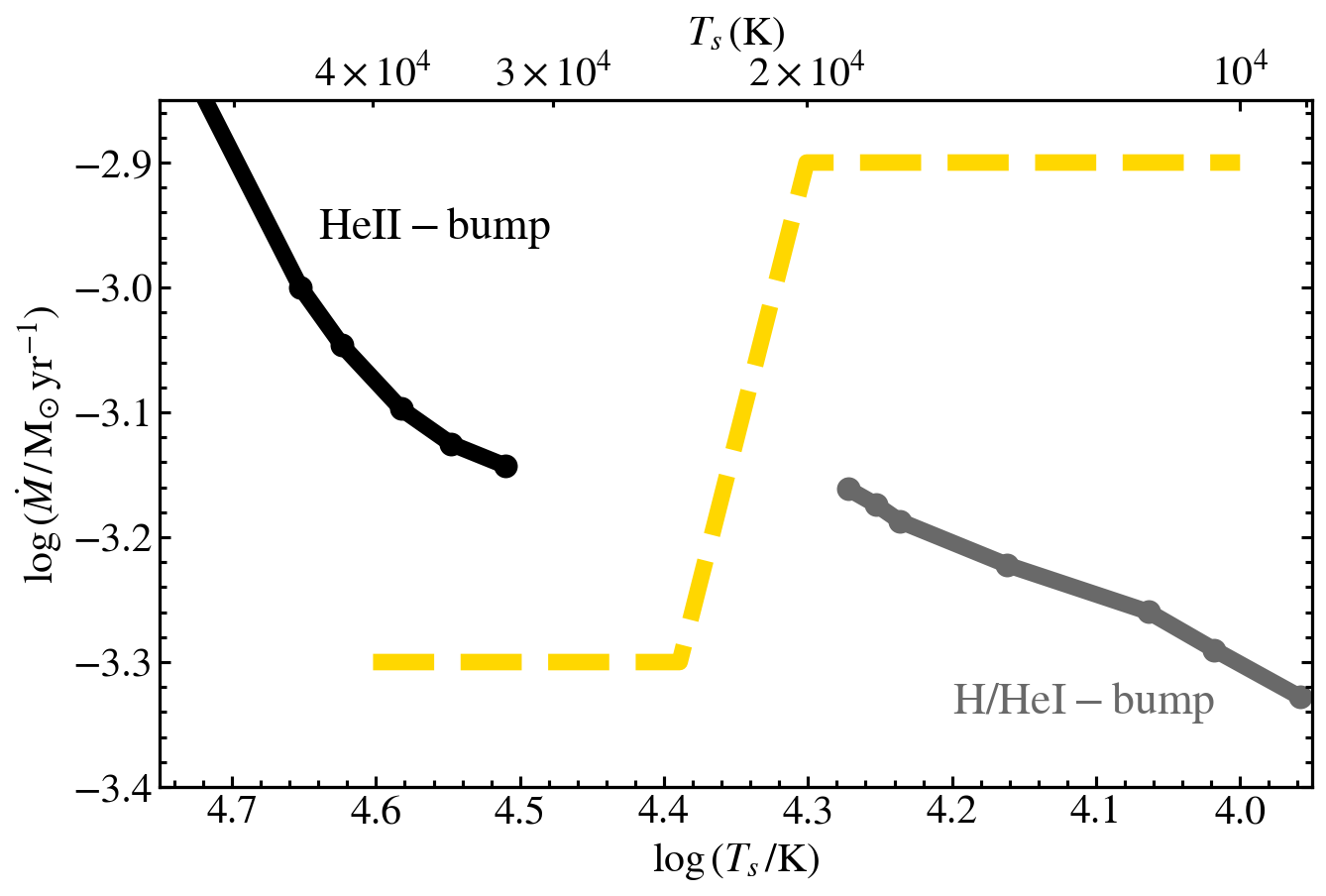}}
\caption{Sonic-point temperatures derived from a set of 73 \Msun stationary hydrodynamic stellar models, adopting several constant mass-loss rates. The sonic points located in the temperature range of the He{\sc ii} opacity bump are shown as solid black, while those located in the temperature range of the H/He{\sc i} opacity bump are shown as solid gray. The two ranges are separated by the forbidden temperature range (Sect.\ref{Sect.steadymodels}). Superposed is the temperature-dependent mass-loss prescription adopted in our time-dependent calculations (dashed gold, see Sect.\ref{Sect.windenvelope}). }
\label{mdottsrel}
\end{figure}

Figure \ref{mdottsrel} shows the loci of sonic-point temperatures associated with different stationary mass-loss rates for a larger set of 73\,\Msun stellar models, including those introduced in Fig.\ref{structure}.
While it is evident that higher mass-loss rates imply higher sonic point temperatures (see also Fig.\ref{structure}), 
a clear dichotomy emerges in Fig. 2, with outflows reaching the sonic points either in the temperature range of the He{\sc ii}-opacity bump for $T_S\gtrsim 30\,$kK or in the temperature range of the H/He{\sc i}-opacity bump for $T_S\lesssim 20\,$kK.  
The possible sonic-point temperatures (and thus sonic-point radii, Fig.\ref{structure}) are discontinuously separated between 30 and 20\,kK. The lack of models in this temperature range can be interpreted considering that in these radiation-pressure-supported envelopes, accelerating an outflow to transonic velocities requires outward-increasing opacities at the sonic point \citep[][]{1999LamersCassinelli,2002Nugis,2018Grassitelli,2019Ro}.
 In other words, at the sonic point, the condition $d\,\kappa/d\,r>0$, or almost equivalently, $\partial\,\kappa/\partial\,T<0$, has to be fulfilled.
This is not the case between 25 and $\rm 30\,kK$, which corresponds to the temperature range where the opacity of stellar matter decreases for decreasing temperatures, after having reached the peak opacity of the He{\sc ii}-opacity bump (Fig.\ref{opacita}).
Within this {\it \textup{forbidden}} temperature range, which lies between the He{\sc ii}- and the H/He{\sc i}-opacity bump,  the outflow in our stellar models exhibits a decrease in radial velocity (Sect.\ref{Sect.appendixmassloss}, see also Sect.\ref{Sect.SonicDiagram}). The forbidden temperature range includes not only the range with decreasing opacities, but also the lower temperature range, where despite the once again increasing opacity, the opacity is still lower than the peak opacity of the hotter He{\sc ii}-opacity bump (i.e. 20-25\,kK, see also Figs.\ref{sonicdiagrammdot}\,\&\,\ref{opacita}). No applied mass-loss rate leads to sonic-point radii in the range $\approx 150-180\Rsun$, as these would correspond to stellar models with sonic-point temperatures in the forbidden range. 

The separation between the solutions at the He{\sc ii}- and at the H/He{\sc i}-opacity bump takes place at a well-defined threshold mass-loss rate, which we call the {\it \textup{helium-minimum mass-loss rate}}. For a given stellar model, the helium-minimum mass-loss rate is the lowest mass-loss rate for which the outflow reaches sonic velocities in the temperature range of the recombination of He{\sc ii} (Fig.\ref{mdottsrel}). Above this threshold mass-loss rate, the sonic point lies in the temperature range with positive opacity gradients associated with the He{\sc ii}-opacity bump. 
Below, the outflow does not accelerate up to sonic velocities in the temperature range of the He{\sc ii}-opacity bump, but such a mass-outflow can eventually reach sonic speed in the temperature range of the more pronounced H/He{\sc i}-opacity bump.
 The threshold mass-flux can be inferred directly from the peak opacity of He{\sc ii}-bump in the OPAL opacity tables, based on the assumption that the radiative and gravitational force equal each other at the sonic point (as we discuss in App.\ref{Sect.SonicDiagram}).

 A characteristic aspect of inflated envelopes at the Eddington limit is that their radial extent is very sensitive to the local balance of forces \citep{2006Petrovic,2011Brott,2012Grafener,2015Sanyal,2017Sanyal}. Fig.\ref{structure} shows that the envelope
structure is significantly altered around the threshold helium-minimum mass-loss rate. This is due to the pivotal contribution of the inertial force in the momentum equation \citep{2006Petrovic} because for nearly transonic flows, it becomes comparable to the gas-pressure gradient (see Appendix \ref{app.inertia}).
The significant impact that different mass-loss rates (and relative inertial forces) around this threshold have on the outer structure of massive star models with inflated envelopes near the Eddington limit implies that both the sonic-point temperature and the sonic-point radius change by a factor 4 owing to the difference in mass-loss rates of a factor 2. 
Both the discontinuous change in sonic-point conditions at temperatures comparable to those at the surface of LBVs and the large variations in the extent of the inflated envelopes due to changes in the mass-loss rates might therefore play a key role in the large variations inherent to the S Dor variability.

\section{Stellar wind-inflated envelope instability}\label{Sect.windenvelope}

Stellar atmosphere models and observations both suggest a rapid increase in the mass-loss rate for decreasing photospheric temperatures at $\approx 25$\,kK of approximately a factor 3 or more \citep[][]{1990Pauldrach,1995Lamers,1999Vink,2001Vink,2002Vink,2004Smith,2016Petrov,2018Vink}. This increase in mass-loss rate, called the bistability mechanism, falls within the forbidden temperature range introduced above. 
Because this increase in mass loss takes place within this well-defined temperature range, a situation may occur without stationary thermal equilibrium configuration for our stellar models (see below).
Motivated by this, we conducted a hydrodynamic stellar evolution simulation to investigate the reaction of the stellar envelope to the mass-loss conditions enforced from a mass-loss prescription that is a simplified representation of the prescription from detailed stellar wind models \citep[][Fig.\ref{mdottsrel}, gold line]{2004Smith}. 

We employed a custom mass-loss relation throughout the bistability temperature (cf. Appendix \ref{Sect.SonicDiagram}) depending on the sonic-point temperature, consisting of two constant mass-loss rates, that is, $\rm log(\Mdot/\Msun\,yr^{-1})= -3.3$ for $\rm T_S>25$\,kK and -2.9 for $\rm T_S<20$\,kK, continuously connected in the range 20--25\,kK (Fig.\ref{mdottsrel}). 
As we show below, for our numerical experiment our focus is on two aspects of the mass-loss prescription. The first is the temperature dependence of the mass-loss prescription, that is, the large increase in mass-loss rate within the forbidden temperature range. 
The second concerns the absolute mass-loss rate values in respect to the threshold helium-minimum mass-loss rate. The threshold mass-loss rate of the adopted stellar model is $\rm log(\Mdot_{He}/\Msun\,yr^{-1})\approx -3.2$ (Fig.\ref{structure}). It is crucial for this numerical experiment that the high and low extremes of the applied mass-loss prescription are set above and below this threshold, respectively. Therefore we uniformly increased the mass-loss prescription by approximately a factor 5 (cf. Fig.\ref{minagcar}) compared to the prescription by \citet{2001Vink} below 20kK (or compared to the estimated mass-loss rates of AG Carinae at maximum brightness, see Sect.\ref{sect.comptobserv}) at the specific luminosity and surface chemical composition of the adopted stellar model. 
Although this might seem a notable artificial increase in mass-loss rate, our aim here is not yet to quantitatively investigate which stars might be subject to instability during their evolution. Rather, this evolutionary model is meant to be a proof-of-concept for the occurrence of an instability.
From it, we can infer the properties and the necessary ingredients of the instability, which can then be used in some preliminary comparison to observations (Sect.\ref{sect.comptobserv}).

\subsection{Evolutionary model}

\begin{figure}
\resizebox{\hsize}{!}{\includegraphics{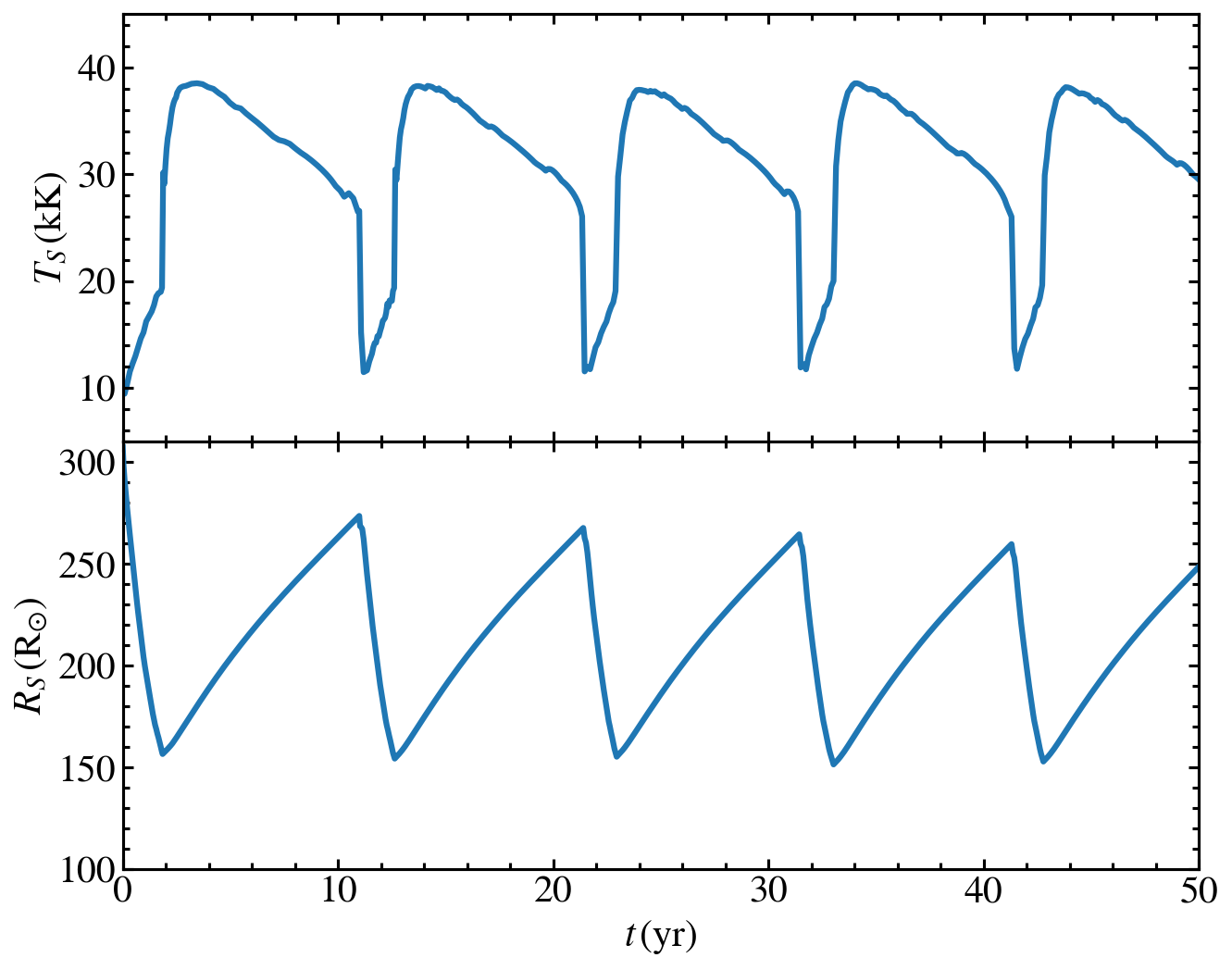}}
\caption{Sonic-point temperature ({\it top}) and radius ({\it bottom}) as a function of time for the first 50 years of the evolution of our time-dependent hydrodynamic stellar evolutionary calculation with sonic-point boundary conditions (Sect.\ref{method}). See also Fig.\ref{RsTs}.
}
\label{tstime}
\end{figure}
We computed the temporal evolution over a few thousand years (the first 50 yr are shown in Fig.\ref{tstime}), starting from the inflated 300 \Rsun stellar model in thermal equilibrium with $\rm T_S\approx 10$\,kK and $\rm log(\Mdot/\Msun\,yr^{-1})\approx -3.3$ in Fig.\ref{structure}.  
In the first two years, the evolutionary model finds itself out of thermal equilibrium as it experiences a higher mass-loss rate, $\rm log(\Mdot_{He}/\Msun\,yr^{-1})= -2.9\,$, larger than the $\rm log(\Mdot_{He}/\Msun\,yr^{-1})= -3.3\,$ of the starting model, as demanded by the imposed mass-loss prescription. The increase in mass-loss rate compared to the initial model implies a higher inertial acceleration, which affects the force balance in the entire envelope, inducing a drastic structural readjustment.
The inflated evolutionary model contracts and deflates, while the sonic-point temperature increases.
The mass-loss rate is above the helium-minimum mass-loss rate, therefore we can infer from Fig.\ref{mdottsrel} that the outflow can reach transonic velocities deeper inside the stellar model, in the proximity of the He{\sc ii}-bump at a much smaller radial coordinate (i.e., $\rm T_S\approx 50$\,kK).
Consequently, the contraction continues as long as the evolutionary models do not readjust and reach an envelope structure consistent with such a mass-loss rate (i.e., the compact models in Fig.\ref{structure}), implying a drastic variation in sonic-point radius.
However, when the sonic-point temperature reaches $\approx 20$\,kK, the now partially deflated stellar model finds itself at the edge of the forbidden temperature range (Fig.\ref{mdottsrel}). A rapid transition takes place, with the sonic point moving from 20 to 30\,kK, that is, from the temperature range of the H/He{\sc i}-bump to that of the He{\sc ii}-bump.
The rapid temperature change at the sonic point of our evolutionary models induces a decrease in the experienced mass-loss rate, as from the mass-loss prescription in Fig.\ref{mdottsrel}.
The contraction halts, with a sonic-point radius approximately half of the initial one. 
 The now lower imposed mass-loss rate is below the helium-minimum mass-loss rates.
Therefore the stellar evolutionary model is not able to sustain the acceleration necessary to a steady transonic outflow in the temperature range of the He{\sc ii}-bump.
Instead, the radiative force in the temperature range of the He{\sc ii}-bump forces the stellar evolutionary model to inflate the subsonic and nonadiabatic stellar envelope once again. The evolutionary models start expanding at a speed of a few \kms, while the sonic-point temperature first reaches a maximum at $\approx\,$40\,kK and then decreases toward the stable sonic-point conditions for this mass-loss rate (i.e., $T_S\approx10\,$kK, Fig.\ref{mdottsrel}). 
At $\approx250\Rsun$, the sonic-point temperature transitions again, this time from $\approx\,$27 to 10\,kK, the applied mass-loss rate increases, the evolutionary models start contracting once more, and the cycle restarts. 
No thermally stable configuration is expected as long as the increase in mass-loss rate takes place within the forbidden temperature range separating the He{\sc ii}- and H/He{\sc i}-bump (Figs.\ref{mdottsrel}\,and \ref{sonicdiagrammdot}). 

Our numerical calculations show a periodic variability with a timescale of $\approx12$ yr, radial amplitudes of $\approx100$\Rsun, and sonic-point temperature variations from 30-40\,kK to 10-20\,kK as the conditions at the base of the radiation-driven stellar outflow periodically change from the He{\sc ii}-bump to the H/He{\sc i}-bump temperature range, without a thermally stable equilibrium configuration (see also Fig.\ref{RsTs}). While in the early phases the evolutionary models are out of equilibrium as a result of the newly imposed mass-loss prescription, the fact that after several Kelvin-Helmholtz timescales the variability of our evolutionary models does not dissipate but becomes almost perfectly periodic shows that the cycle is self-sustained. 
This lack of a stable configuration can be understood from Fig.\ref{mdottsrel}, where the stable sonic-point conditions are plotted together with the mass-loss prescription adopted in our evolutionary models. The mass-loss prescription never crosses the two separate loci of thermally stable solutions associated with the He{\sc ii}- and the H-bump. This implies that our evolutionary model is unable to find a stable configuration in thermal equilibrium, and thus the persistence of the cycle. 
After several hundreds years, the evolutionary models thermally relaxes to a well-defined periodic cycle, regardless of the initial conditions. 
In our numerical models the cycle arises naturally from the setup of the simulation, without the need of external or internal trigger other than meeting the conditions that are required for the instability, that is, a massive inflated stellar model near the Eddington limit, the appearance of a forbidden temperature range, and a mass-loss prescription that crosses the helium-minimum mass-loss rate within this temperature range.
The opposite is true for evolutionary models that do not meet these conditions (i.e., do not have a mass-loss prescription that crosses the helium-minimum mass-loss rate of this stellar model, do not present a forbidden temperature range, or do not develop inflated envelopes due to the proximity to the Eddington limit).

The $\approx10$ yr period is consistent with the thermal timescale of the inflated envelope, which starts from $\rm log(T/K)\approx6.4$ \citep[see the criterion in Appendix A by][]{2018Grassitelli}, that is, from the iron L-shell recombination temperature. The dynamical timescale of the stellar envelope is on the order of a week. The low density and small heat capacity within these inflated envelopes imply that nonadiabatic effects are important, with local energy losses and gains per mass element due to the structural readjustments.
During the contraction and the expansion phase, the evolutionary models redistribute $\approx 0.01-0.1$\Msun of material and its energy content. Considering that the thermal structure is coupled to the hydrodynamics of our stellar models at the sonic point, the thermal readjustment of the whole inflated envelope sets the timescale for the cycle.

\subsection{Sonic point radius-temperature cycle}\label{sect.rsts}

\begin{figure}
\resizebox{\hsize}{!}{\includegraphics{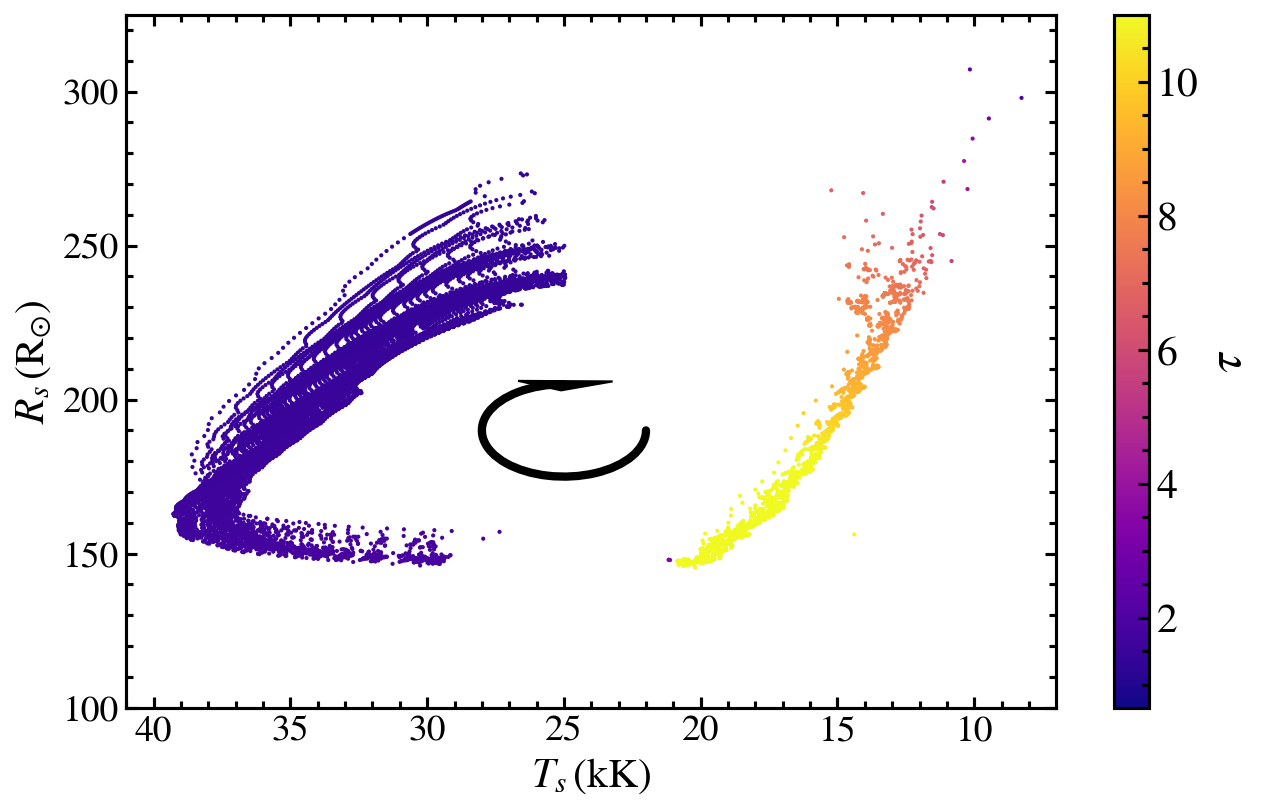}}
\caption{Sonic-point radius and temperature during the $\approx 3*10^3\,$yr of evolution of our numerical calculations (Sect.\ref{Sect.windenvelope} and Fig.\ref{tstime}). Each point corresponds to a model ($\approx$100000 models), color-coded as a function of the estimated optical depth at the sonic point (color bar on the right side). The S Dor cycle proceeds clockwise, as indicated by the black arrow.}
\label{RsTs}
\end{figure}

Figure \ref{RsTs} shows the sonic-point temperatures and radii of our evolutionary calculation in Fig.\ref{tstime} over the whole $\approx\,$3000 yr of our numerical simulation. The large number of semi-regular cycles undergone in our calculations periodically cover a parameter space from 40 to 10\,kK and from 150 to 280\,\Rsun. 
It can be seen how, as the evolutionary model thermally relaxes and reaches periodicity, the parameters of the models computed during the cycles partially change, but remain confined to an almost circular range of sonic-point radii and temperatures (similar to those displayed in Fig.\ref{tstime}).
Most of our computed models populate the temperature range with $T_s>25\,$kK, manifesting a clear scarcity of models within the forbidden temperature range. This indicates that the transition between configurations with sonic point in the temperature range of the He{\sc ii}- and H-opacity bump takes place on a dynamical timescale that is shorter than the adopted time step. 
Because the mass-loss prescription does not contemplate a discontinuous change as a function of temperature, an intermediate configuration would have been possible in principle. This means that the lack of an equilibrium configuration is just due to the inability of our massive stellar models to develop a stationary transonic outflow within the forbidden temperature range (Fig.\ref{mdottsrel}). 

Figure \ref{RsTs} also indicates the Rosseland optical depth of the wind at the sonic point of our stellar models, estimated with the simplified wind models introduced in Sect.\ref{method}.
The hot ($T_s>25\,$kK) stellar models have low optical depths ($\approx 1$), while cool stellar models have optical depths of about 10. The formation of an optically thick wind for the cool ($T_s<20\,$kK) stellar models extends the photospheric radius by approximately a factor 2 compared to the sonic-point radii.

\section{Comparison with observations}\label{sect.comptobserv}

\begin{figure}
\resizebox{\hsize}{!}{\includegraphics{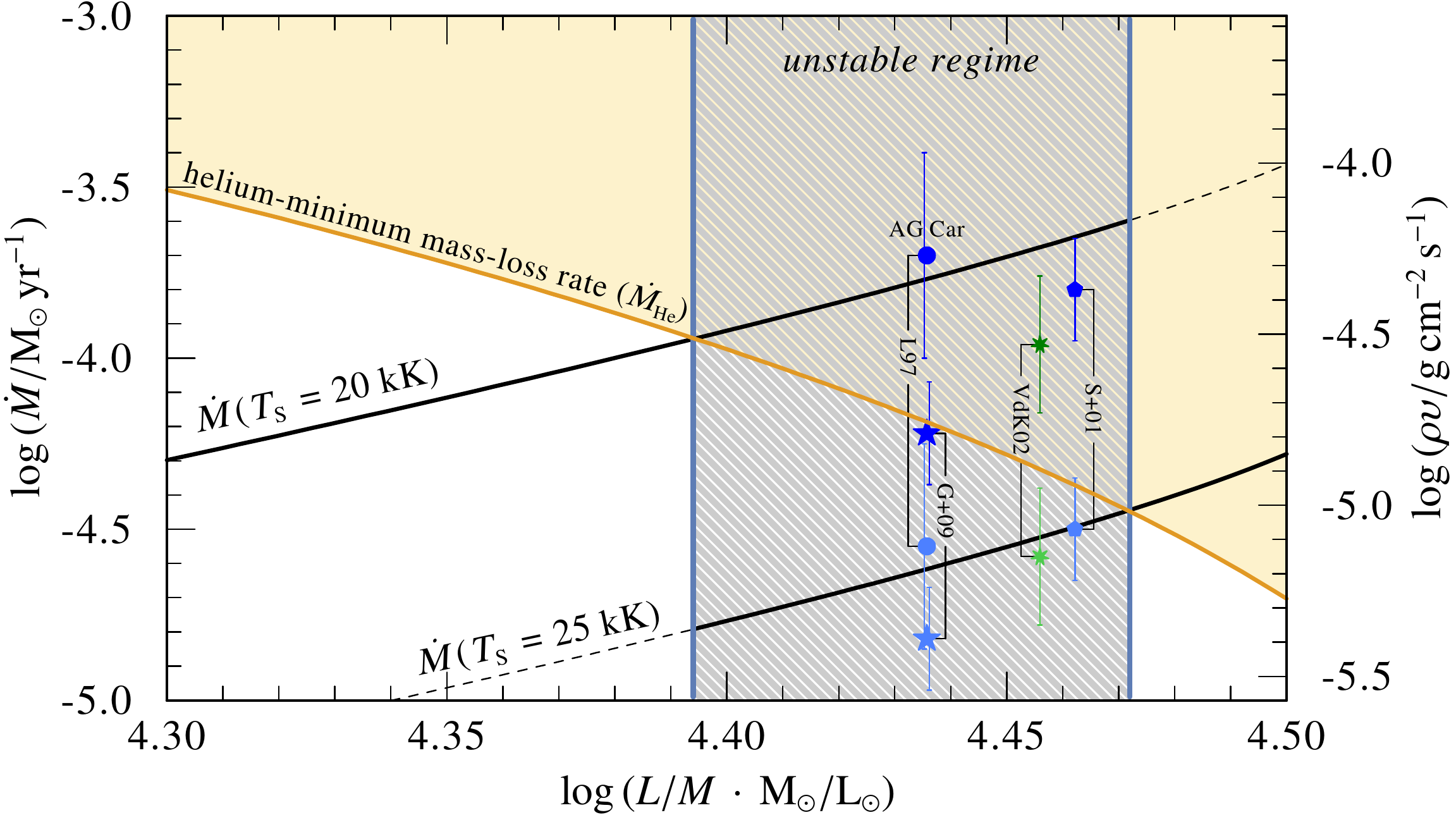}}
\caption{Minimum mass-loss rate (left axis) and minimum mass-flux (right axis) for the sonic point of a radiation-driven outflow to lie in the temperature range of the He{\sc ii}-opacity bump (thick orange line) as a function of L/M. The helium-minimum mass-loss rate is derived based on the opacity tables (see Appendices \ref{Sect.SonicDiagram} and \ref{sect.opacita}) adopting a sonic-point radius of 62\Rsun \citep{2011Groh}, Y=0.6 \citep{2009Groh}, and a turbulent velocity broadening of 20\,\kms \citep{2011Groh}. The luminosity and mass adopted for AG Car are $1.5\cdot10^6$\Lsun and 55\Msun \citep{2011Groh,2015Vamvatira}. The blue points indicate the minimum and maximum mass-loss rates of AG Carinae from \citet[][stars]{2011Groh}, \citet[][circles]{1994Leitherer}, and \citet[][pentagons]{2001Stahl}, as well as the numerical predictions specific to AG Carinae by \citet[][green asterisks]{2002Vink}. The black lines indicate the mass-loss rate predictions on the hot (25\,kK, lower line) and cool (20\,kK, upper line) side of the bistability temperature \citep{2001Vink}, derived adopting a mass-luminosity relation \citep{2011Grafener}. The gray area indicates the L/M range where we predict massive main-sequence stars might undergo an S Dor phase (cf. Appendix \ref{sect.initmassrange}).}
\label{minagcar}
\end{figure}

Our scenario and our proof-of-concept evolutionary model suggest that the radiative force in the temperature range of different opacity bumps is responsible for the early wind acceleration during minimum and maximum brightness phase of an S Dor cycle.
Therefore we can postulate that LBVs experiencing S Dor variations have mass-loss rates that cyclically fall above and below their threshold helium-minimum mass-loss rate. 
Considering that the threshold mass-flux can be directly derived from the adopted opacity tables (see Appendix \ref{Sect.SonicDiagram}), this aspect is independent of the detailed characteristics of individual evolutionary models and provides an easy way to test our scenario on an observational basis: we expect the minimum and maximum brightness phase to be on opposite sides of the helium-minimum mass-loss rate.

Figure \ref{minagcar} shows the helium-minimum mass-loss rate and mass-flux as a function of L/M, derived based on the physical parameters (i.e., radius and surface abundances) of the archetypal LBV AG Carinae. The helium-minimum mass-loss rate in Fig.\ref{minagcar} does not correspond to that of the model in Sect.\ref{Sect.windenvelope}, mostly because different radii (62 \Rsun and 150 \Rsun, respectively) and surface chemical abundances are considered. We opted for AG Carinae because of the Galactic LBVs undergoing an S Dor cycle, it is the best studied \citep{1988Lamers,1989Lamers,1997Crowther,2001Stahl,2009Groh}, and its luminosity and surface chemical composition is very similar but not identical to that of our evolutionary models in Fig.\ref{tstime}. Moreover, it is one of the very few LBVs for which stellar wind models at different phases of the S Dor cycle are available \citep{1994Leitherer,1997Leitherer,2001Stahl,2002Vink,2011Groh}. 
However, only \citet{2001Stahl} and \citet{2002Vink} provided the mass-loss rates during almost a full cycle. \citet{2011Groh} instead computed atmosphere models near the bistability temperature (i.e., from 24 to 14\,kK), showing how the mass-loss rates rapidly increase for decreasing temperatures, while \citet{1997Leitherer} only provides mass-loss rates at minimum and in the early stages of the approach of the AG Car maximum brightness phase.

Figure \ref{minagcar} shows that the highest and lowest empirical mass-loss rates reported for AG Carinae are found on different sides of the helium-minimum mass-loss rate, that is, the maximum mass-loss rate is above and the minimum is below, supporting our scenario. The mass-loss rates of AG Car are systematically higher in the cool maximum brightness phase, and lower in the hot minimum brightness phase (see \citealt{2001Stahl} and fig.5 by \citealt[][]{2002Vink}). Moreover, in Fig.\ref{minagcar}, it is important to notice that the helium-minimum mass-loss rate rapidly decreases for increasing L/M, suggesting that stars with high L/M require lower mass-loss rates to cross this threshold and potentially trigger the S Dor cycle (see Appendix \ref{sect.initmassrange} for an estimate of the initial mass range of core-H-burning stars undergoing S Dor cycles). 
This also suggests that more luminous stars would more easily reach the threshold mass-loss rate at higher sonic-point temperatures, potentially accounting for the temperature dependence in the distribution of the S Dor variables in the HR diagram \citep{2012Vink}.
The results in Fig.\ref{minagcar} are independent of the evolutionary models in Sect.\ref{Sect.windenvelope} because they exclusively rely on the atomic physics from the adopted opacity tables that is used to derive the helium-minimum mass-loss rate, and upon the empirical estimates of the AG Carinae mass-loss rates. 
Despite the apparent agreement of our theoretical expectations and the AG Car observed mass-loss rates, a word of caution is necessary. The observational values and our theoretical threshold mass-loss rate are both uncertain. We therefore consider it to be already remarkable that this pioneering investigation finds comparable orders of magnitude. 

\begin{figure}
\resizebox{\hsize}{!}{\includegraphics{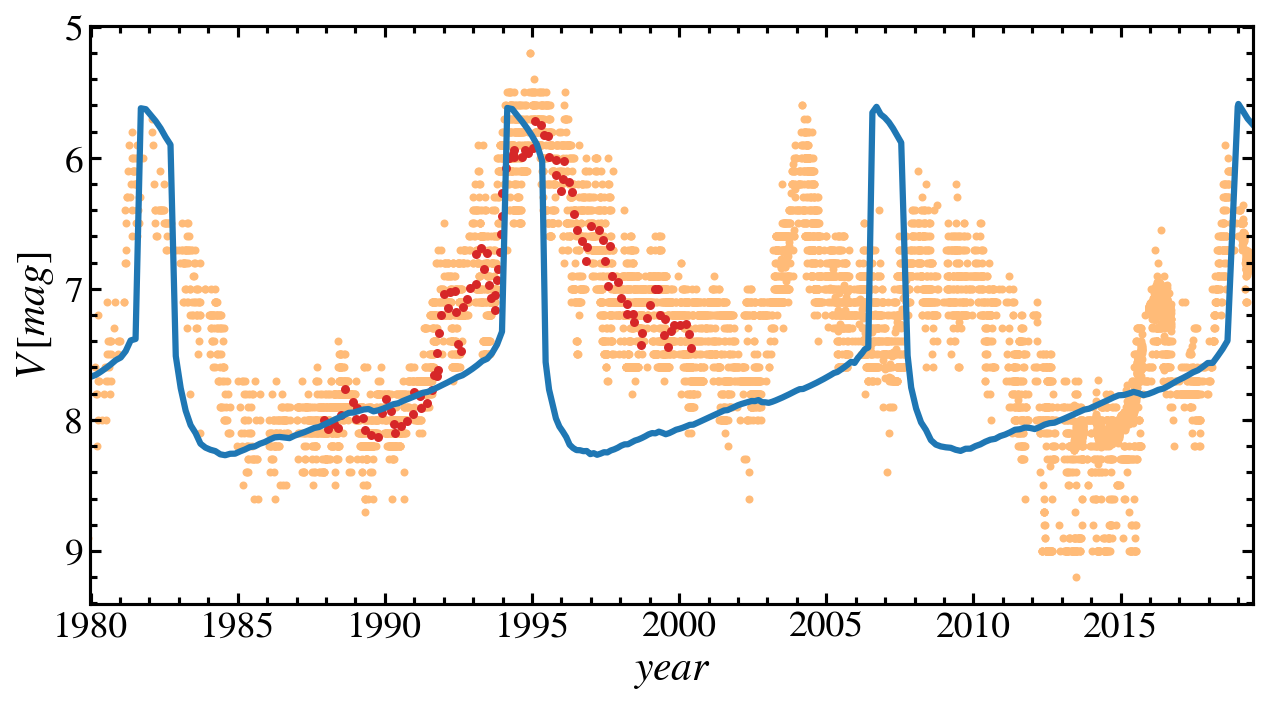}}
\caption{Visual magnitude as a function of time for AG Car from Stahl et al.(2001, red) and AAVSO (www.aavso.org, orange) between 1980 and 2019. Superposed, we show the visual magnitude variations associated with the photospheric conditions of our hydrodynamic stellar evolution calculation, derived with our simplified stellar wind models (blue, Sect.\ref{method}).
}
\label{vismag}
\end{figure}

We then compared the photospheric brightness evolution expected from our stellar evolution and wind calculations to those observed in the LBV AG Carinae \citep{2001Stahl} between 1980 and 2019 (Fig.\ref{vismag}). 
Our brightness profile shows a regular increase of approximately 2 magnitudes on a timescale of $\approx 12\,$yr. The rapid increases in brightness mainly take place when the sonic-point temperature transitions from the He{\sc ii}-bump to the cooler H-bump temperature range, which also leads to the formation of a denser optically thick wind, and vice versa (Fig.\ref{RsTs}). The maximum brightness phase lasts for a few years, corresponding to the stage in which the sonic point is in the temperature range of the H-bump.  

\begin{figure}
\resizebox{\hsize}{!}{\includegraphics{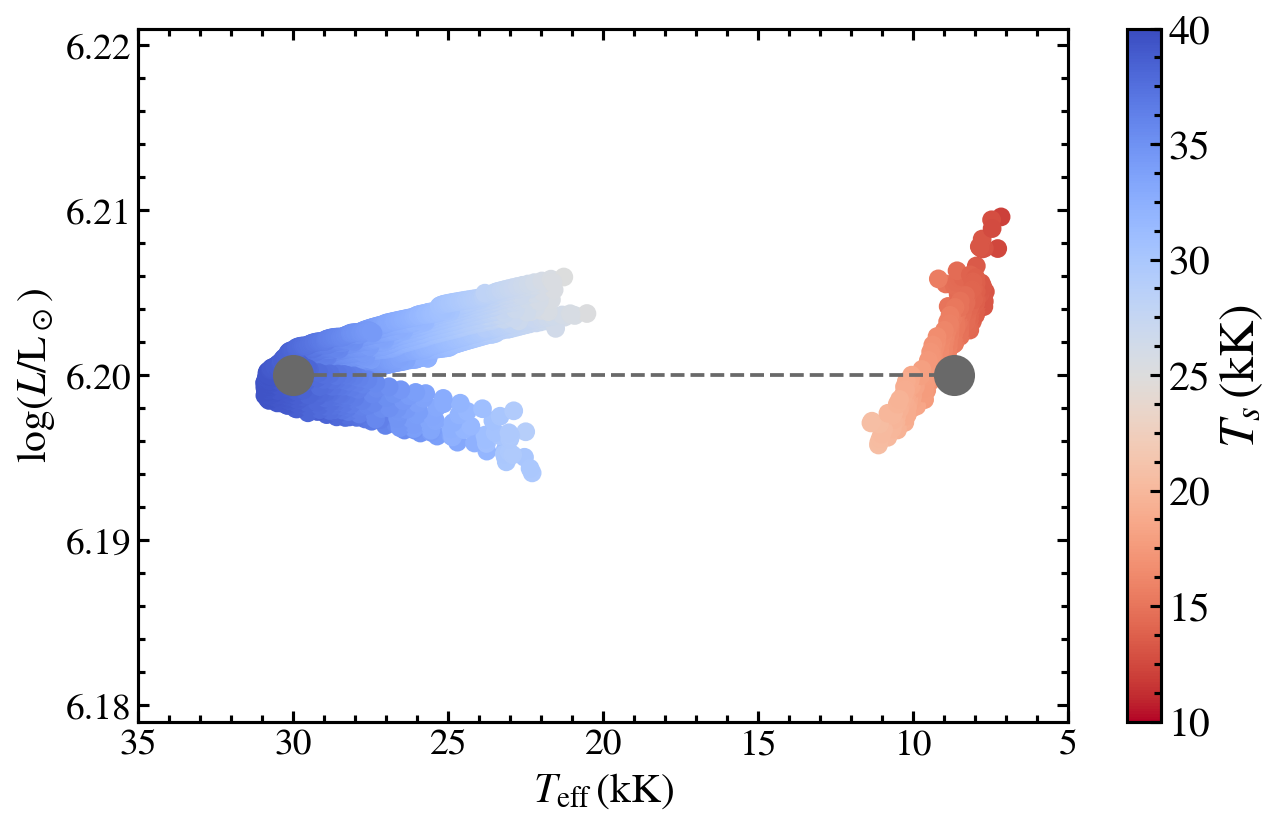}}
\caption{Several cycles from our evolutionary models in Sect.\ref{Sect.windenvelope} in the HR diagram, color-coded according to their sonic-point temperature. Each dot corresponds to a computed stellar model during our evolutionary calculation. Superposed, we show the location in the HR diagram of the minimum and maximum brightness phase of AG Carinae (gray dots and dashed line) from \citet[][fig.1]{2012Vink}. The effective temperature at optical depth unity is estimated with the wind models discussed in Sect.\ref{method}, while the bolometric luminosity has been decreased by 0.05 dex to correspond to that of AG Carinae.}
\label{HRD}
\end{figure}

 Figure \ref{HRD} compares the location in the HR diagram of AG Carinae \citep[as in fig.1 by][]{2012Vink} to the photospheric conditions of some of our evolutionary models described in Sect.\ref{Sect.windenvelope}. The effective temperature at optical depth unity was estimated with the simplified stellar wind models discussed in Sect.\ref{method}, and is only considered a rough estimate. Nonetheless, when we compare the AG Car observed temperature variations to those of our stellar models, the similarities are encouraging.  During the hot phase of the cycle, the effective temperature is restricted to the range $25-30\,$kK, having low optical depths at the sonic point.
In the cool maximum brightness phase, the effective temperature falls within the range $7-11\,$kK (Fig. \ref{HRD}), corresponding to optical depths of about 10.
This is not far from the results of stellar wind models by \citet{2011Groh}, which show that the Lyman continuum is thin or partially optically thin for temperatures higher than $\approx 20\,$kK, while it becomes severely optically thick (i.e., $\tau_{Lyc}\approx 10$) at temperatures below $\approx 20\,$kK.      
The bolometric luminosity at the sonic point shows changes of about 1\% during the different phases of the cycle as part of the luminosity is transformed into work for the expanding or contracting envelope. We find no systematic luminosity variation between the hot and cool phases of the cycle. However, the variable wind kinematics might affect the photospheric luminosity during the cycle.

Despite the differences between our evolutionary model, meant to be a proof-of-concept of the wind-envelope instability, and some parameters of AG Car, such as its large rotational velocities, the amplitude of the brightness variations, the locations in the HR diagram, and the average period agree well with observations. The duration and ratio of the maximum and minimum brightness ($\approx\,$2 and 9 years, respectively) also agree fairly well with the AG Car light curve. 
However, our 1D evolutionary models are unable to reproduce the apparently stochastic microvariability on the dynamical timescale, and the variations in period and rise or decline time between peaks (see Sect.\ref{DiscussConcl}). 
Although the observed brightness variations of AG Carinae in the last 40 years appear partly irregular, and our evolutionary calculations only partially reproduce features in the AG Car light curve, the overall phenomenology of our models for the first time resembles the quasi-periodic large brightness variations typical of LBVs experiencing S Dor variations.

\section{Discussion and conclusion}\label{DiscussConcl}

Our study focused on the hydrodynamic treatment of stellar envelopes and transonic outflows.
It suggests that the S Dor variability of LBVs may arise due to the nonmonotonic temperature-dependent form of the local Eddington limit in the outer stellar layers, and the variable conditions at the base of their dense radiation-driven winds \citep[morphologically consistent with the contours representing the Eddington limit derived through model atmospheres by][]{1988Lamers,1995Nieu}. 
We find that when a stellar model crosses the well-defined threshold helium-minimum mass-loss rate, a rapid change in the conditions at the base of the wind takes place, with the sonic point moving between the temperature ranges of the He{\sc ii}- and H/He{\sc i}-opacity bump. This leads to dramatic structural changes in the stellar envelope that in turn cause mass-loss variations that reverses the initial change. 
A cycle arises because of the lack of a stable equilibrium configuration.
Three key physical ingredients are required to trigger the instability: close proximity to the Eddington limit, leading to the formation of inflated stellar envelopes and radiation-driven dense stellar outflows;  temperature ranges with decreasing opacities, lacking the increase in radiative force relative to the local gravitational force, which is required to develop transonic stellar outflows; and an increase in mass-loss rate that crosses the threshold helium-minimum mass-loss rate within the forbidden temperature range with decreasing opacities, implying the lack of a stable envelope configuration.

 High L/M prominently appear especially during the late phases of stellar evolution, as stars naturally increase the L/M during their lifetime \citep[e.g.,][]{2011Brott}. 
The proximity to the Eddington limit also drastically increases the mass-loss rate, and when the conditions are such that a star can cross the threshold helium-minimum mass-loss rate within the forbidden temperature range, we might expect the onset of this self-sustained cycle.

In our evolutionary test case presented in Sect.\ref{Sect.windenvelope}, the helium-minimum mass-loss was rather high. In order to exceed it, we had to introduce mass-loss rates that were approximately five times higher than observed for AG Car.
While this is not ideal, numerous uncertainties affect the absolute values involved in our numerical experiment, which could mitigate the discrepancy.  
In primis, the threshold mass-loss rate is mostly determined by the L/M of the stellar model and can rapidly decrease as the L/M increases.

It remains to be seen whether our qualitative scenario is able to quantitatively reproduce the detailed observational properties of the numerous LBVs undergoing an S Dor cycle.
From this point of view, a large set of hydrodynamic evolutionary calculations is required to investigate if and when the conditions to trigger the cycle are met during the evolution of stellar models. Only such a comprehensive analysis would be able to show whether the luminosity-to-mass ratios and the state-of-the-art mass-loss prescriptions of blue and yellow supergiants allow the S Dor variability, as suggested by our scenario. The comparison to AG Car (Fig.\ref{minagcar}) already suggests that the observed mass-loss rates are comparable to the expected threshold. However, reproducing the distinct observational distribution of S Dor variables in the HR diagram and their properties might require a demanding scientific effort. This is an opportunity to better constrain the conditions and physics in the envelopes and winds of these extreme stars at the Eddington limit.

Luminous blue variables appear as a heterogeneous group of unstable stars from different stellar environments, potentially in different phases of evolution, and with a variety of light curves \citep{1988Lamers,1989Lamers,1994Humphreys,1997Davidson,2001vanGenderen}. 
In addition to the specific parameters and the threshold helium-minimum mass-loss rate of the stellar model, the characteristics of the cycle appear to be rather sensitive to the imposed mass-loss prescription and the conditions in the inflated envelope. We thus expect that not only the mass and luminosity, but also the temperature-dependent mass-loss prescription, the surface chemical composition, the envelope configuration, etc. all significantly affect the brightness variations of LBVs. 
Uncertainties also involve the treatment of convection, turbulent line-broadening, clumping, and rotation on the structure and the force balance in such low-density envelopes. Especially fast rotation \citep{2020Zhao} and turbulent convection \citep{2020Schultz} could contribute to reaching the local Eddington limit in stellar envelopes. The outward-directed centrifugal force or the higher radiative force from the opacity enhancement due to turbulence can contribute to accelerating the outflow to transonic velocities and in this way lower the threshold mass-loss rate to trigger the instability (potentially compensating for the artificial increase in mass-loss rate in our proof-of-concept evolutionary model). It is also possible that the Rosseland mean opacity underestimates the flux-mean opacity in the dense and inhomogeneous atmosphere of these massive stars, which would also reduce the threshold to trigger the instability. We have neglected these effects because they are potentially important to explain the asymmetry in the LBV ejecta nebulae \citep{1995Nota}, for example, but they are not essential to this novel physical scenario.

The often-quoted theoretical candidate to explain the eruption events in LBVs is the so-called geyser model \citep{1992Maeder}. This qualitative phenomenological model relies upon the idea that density inversions near the surface are washed out by local dynamical instabilities, inducing an ionization front in the stellar envelope that is hypothetically able to eject a significant fraction of the outer stellar layers. 
Our numerically and theoretically supported scenario instead refers to the more ubiquitous S Dor variables and is based on the interplay between low-density inflated envelopes at the Eddington limit and dense radiation-driven stellar winds. The numerical results of our simulations appear to be very promising because they are able to reproduce the most emblematic features associated with the S Dor variability. However, at this stage, our evolutionary models do not lead to eruptive events that could resemble the eruption of $\eta$-Carinae (or similar objects, if any).

Luminous blue variables are often indicated as direct supernova progenitors \citep{2006Kotak,2007Pastorello,2009GalYam,2011Smith,2013Groh,2018Boian,2018Elias,2020Bruch}, which might seem in contrast with our core-H-burning stellar models.
However, in our scenario, the S Dor cycle is not directly linked to the conditions in the core, depending only on the physical conditions within their inflated envelopes.   
Consequently, LBV properties can appear in different evolutionary phases, even several times during a single stellar evolution (including LBV properties for classical WR stars, see Appendix \ref{sect.initmassrange}).

For a thermal timescale variability of several years, the inflated envelopes need to redistribute a fraction of a solar mass.
The results from stellar models with gray atmosphere boundary conditions by \citet{2015Sanyal} showed that only massive stellar models with high L/M in their late main-sequence (or post main-sequence) evolution are close enough to the Eddington limit to develop sufficiently massive inflated envelopes. These models are located in the top right corner of the HR diagram at effective temperature below 20\,kK, and would correspond to red or yellow supergiants.
We can speculate that the S Dor instability with timescales of several years appears in stars that, if it were not for their unstable envelopes and high mass-loss rates, would populate lower temperatures (i.e., $\approx 10-20\,$kK). The instability itself and the high mass-loss rates might thus prevent inflated massive stars from populating the top right corner of the HR diagram \citep{1994Humphreys}, keeping them more compact and hot, and allowing for excursions to lower temperatures only as part of the cycle.
                                                                                          
Nevertheless, we are still far from completely understanding the phenomenon, and detailed hydrodynamic stellar evolution calculations are required to constrain quantitatively which stars undergo an LBV phase, and the effect that this phase has on the evolution of massive stars and their final fate. Moreover, multidimensional calculations including a local treatment of the radiative acceleration in the stellar envelope, and hydrodynamic wind models able to investigate the complex radiative acceleration in the supersonic outflows \citep{2019Sundqvist,2020Sander,2020Bjorklund}, can give insights into the ability of our scenario to reproduce the observed phenomenology of these massive stars.
Furthermore, our models establish the framework required for quantitative predictions on the formation of circumstellar nebulae surrounding LBVs \citep{1995Nota,2003Weis,2014Agliozzo}, due to wind-wind interaction and variable mass-loss rates \citep{2002Vink,2020Koumpia}, as well as their imprint on supernovae \citep{1994Chevalier,2006Kotak,2018Boian}.

This work emphasizes a paradigm shift in our understanding of the properties of massive stars. Mass loss by stellar wind does not only significantly affect the position of stars in the uppermost part of the HR diagram by affecting their mass and chemical composition. It also significantly affects the outer envelope structure \citep{2006Petrovic,2018Grassitelli}, potentially triggering variability when in the proximity of the Eddington limit. 
Stellar evolutionary calculations neglecting a hydrodynamic treatment of the outer stellar layers and the effect of strong stellar winds on stellar envelopes are unable to reproduce this variability and might misrepresent the distribution and evolution of massive stars. 
The relative uncertainties in terms of stellar radius and temperature might, among other things, induce significant uncertainties in close binary systems undergoing mass transfer due to Roche-lobe overflow, precursor of double black-hole systems, or gamma-ray bursts \citep{2016Levan,2016Marchant,2016Mandel,2020Langer}.

\begin{acknowledgements} 
G. G., N.J. G., A.A.C. S., and J.S. V. all contributed equally to this manuscript.\\
L.G. thanks the referee, A. Istrate, J. Bestenlehner, and A. Schootemeijer.  G.G. thanks the Deutsche Forschunsgemeinschaft (DFG) for financial support under grant No. GR 1717/5-1. J.M. acknowledges funding from a Royal Society-Science Foundation Ireland University Research Fellowship (14/RS-URF/3219). J.S. V. and A.A.C. S. are supported by STFC funding under grant number ST/R000565/1.

\end{acknowledgements} 
\bibliographystyle{aa}
\bibliography{sonic}
\appendix

\section{Effect of the inertial acceleration on the outer stellar structure}\label{app.inertia}
The momentum equation in our calculation is
\begin{equation}\label{momentumeq}
 \left(\frac{a}{4 \pi r^2}\right)_t = -\frac{G m}{4 \pi r^4} - \frac{\partial P_{gas}}{\partial m} - \frac{\partial P_{rad}}{\partial m}\quad,
\end{equation}
where $m$ is the mass coordinate, $r$ is the radial coordinate, and $a$ is the inertial acceleration, which in a steady-state condition becomes $a = \upsilon\, \partial \upsilon / \partial r$. 
In the inflated envelopes of massive stars, where convective energy transport is very inefficient, the radiation pressure gradient can be written as
\begin{equation}\label{radiationpressuregradient}
\frac{\partial P_{rad}}{\partial m} = -\frac{\kappa L}{16 \pi^2 r^4 c} = -\frac{g_{rad}}{4 \pi r^2} \quad,
\end{equation}
where $g_{rad}$ is the radiative acceleration, $L$ is the luminosity, $c$ is the light speed, $\kappa$ is the frequency-independent local opacity of stellar matter, and where we have made use of the diffusive approximation for radiative transfer.
Envelope inflation appears in response to a steep increase in opacity in radiation-pressure-dominated layers close to the Eddington limit. In the proximity of the Eddington limit, the radiation pressure gradient almost completely counterbalances the gravitational force, without the need of steep gas pressure gradients in Eq.\ref{momentumeq}. This leads to the characteristic low-density extended envelope with high gas pressure scale-heights, as in Fig.\ref{structure} \citep{2017Sanyal,2018Grassitelli}. However, for stationary transonic flows, that is, $\upsilon \approx c_s$, the usually negligible inertia becomes comparable to the gas pressure gradient, and almost exactly equals it at the sonic point,
\begin{equation}
\frac{\partial P_{gas}}{\rho \,\partial r}= \frac{\partial \rho  c_s^2}{\rho \,\partial r}\approx 2 c_s \frac{\partial c_s}{\partial r} - \frac{c_s^2}{\upsilon} \frac{\partial \upsilon}{\partial r}
 = c_s \frac{\partial c_s}{\partial r}=\upsilon \frac{\partial \upsilon}{\partial r}\quad,
\end{equation} 
where we have made use of the differential form of the continuity equation and neglected the geometrical factor.
It is therefore expexted that the greatest variations in the radial extent appear in the proximity of the helium- (or iron-) minimum mass-loss rate, where the flow velocities within the stellar envelope become transonic \citep{2006Petrovic,2018Grassitelli}.

 The radial extent of inflated envelopes and the sonic-point radii of our stellar models are also affected by the still poorly constrained mixing length parameter and stellar rotation. Lower mixing length parameters lead to more inefficient convective energy transport, hence to more radially extended inflated envelopes \citep{2015Sanyal}, and likely to larger radial variations during the cycle.
Instead, rotating stars experience the outward-directed centrifugal force. They are expected not only to develop more extended and latitude-dependent inflated envelopes \citep{2011Brott,2015Sanyal,2017Sanyal}, but also to have a lower helium-minimum mass-loss rate than the same nonrotating stars. This is due to the acceleration in addition to the radiative acceleration that is provided by the centrifugal force.
Therefore fast-rotating supergiants might trigger the wind-envelope instability for lower mass-loss rates.

\section{Sonic-point diagram}\label{Sect.SonicDiagram}
\begin{figure}
\resizebox{\hsize}{!}{\includegraphics{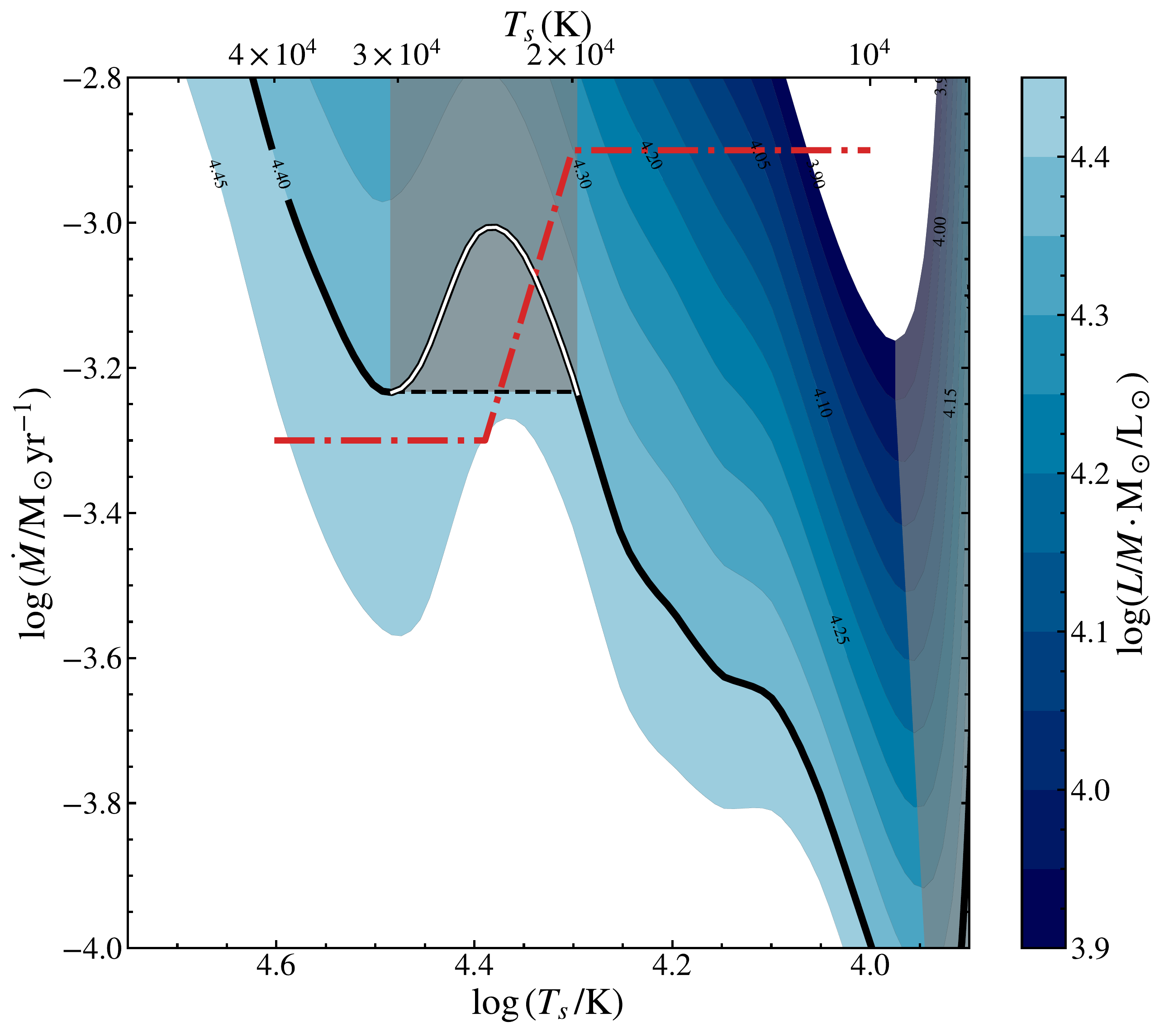}}
\caption{Sonic-point diagram associating the sonic-point temperature and the mass-loss rate to the contours of luminosity-to-mass ratio (for a metallicity 0.02 and a helium mass-fraction 0.48), color-coded according to the bar on the right side. The thick black line follows the contour indicating the possible sonic-point conditions of a star with $\rm L/M =10^{4.4} \Lsun/\Msun$. The white section of this contour and the gray shaded areas indicate the forbidden temperature range. The dot-dashed red line indicates the mass-loss prescription adopted in our numerical calculations. }

\label{sonicdiagrammdot}
\end{figure}

The Eddington limit is a limit on the hydrostatic stability of stars \citep{1997Langer,2012Langer,2015Sanyal}. 
The proximity to this limit can be defined locally by the Eddington factor $\Gamma$, given by
\begin{equation}\label{gamma}
\Gamma=\frac{g_{rad}}{g}=\frac{\kappa L}{4 \pi c G M}
,\end{equation}
where $g$ is the gravity, $M$ is the stellar mass, and $G$ is the gravitational constant. 
For radiation-driven stellar winds as well as our stellar models in Fig.\ref{Sect.steadymodels}, the sonic point is located at $\Gamma\approx1$ \citep{1999LamersCassinelli,2002Nugis,2018Grassitelli,2020Sander}. We can therefore define the Eddington opacity $\kappa_{Edd}$ as
\begin{equation}\label{kappaedd}
\kappa_{Edd}=\frac{4 \pi c G M}{L}\quad.
\end{equation} 
The Eddington opacity is the opacity at the sonic point of a radiation-driven wind, which is solely determined by the luminosity-to-mass ratio. 

It is possible to directly relate the wind density to the sonic-point temperature at which a star finds sufficient radiative force to overcome the gravitational pull and launch a supersonic stellar outflow according to the commonly used opacity tables (which for a given chemical composition, associate the Rosseland opacity of stellar matter with a density and temperature, Fig.\ref{opacita}). We do so using the so-called sonic-point diagram \citep[Fig.\ref{sonicdiagrammdot},][]{2018Grassitelli}. This diagram directly relates the mass-flux by stellar wind at the sonic point (i.e., $\rho\cdot\upsilon$ or, at the sonic point, $\rho\cdot c_s$, where $c_s$ is the sound speed) to the temperature of the sonic point for a given luminosity-to-mass ratio. 
The sonic-point diagram in Fig.\ref{sonicdiagrammdot} is derived and shaped by the commonly used OPAL opacity tables \citep{1996Iglesias}, here adjusted for the effect of turbulent broadening of the spectral lines (see below). It directly associates an L/M with the contours of opacity from the OPAL tables (Eq.\ref{kappaedd}). The density can then be replaced by its product with the isothermal sound speed (function of temperature and chemical composition only), that is, the mass flux.
In this way, we can define the loci of possible sonic-point conditions of a star with a purely radiation-driven wind (Fig.\ref{sonicdiagrammdot}), without the need to compute large sets of hydrodynamic stellar models as in Figs.\ref{structure}\,and\,\ref{mdottsrel}. Then, mass fluxes can be converted into more familiar mass-loss rates with the adoption of a sonic-point radius. This implies, however, that the sonic-point diagram has to vary when it is applied to different stellar models with different stellar parameters (as noted for Fig.\ref{minagcar}). For example, the quoted mass-loss rates and contours quadratically depend upon the chosen reference radius.
  
The sonic-point diagram can be used, given an L/M and a mass-loss rate, to infer the sonic-point conditions of purely radiation-driven stellar winds, under the assumption that the Rosseland mean opacity is representative of the flux-weighted opacity at the sonic point. For example, from Fig.\ref{sonicdiagrammdot}, a star of 150\Rsun and $\rm L/M=10^{4.4}\Lsun/\Msun$ (as in Figs.\ref{structure}\,and\,\ref{mdottsrel}) that loses mass at a rate of $\rm \Mdot=10^{-3}\Msun/yr$, that is, with a very dense and optically thick wind, is expected to launch its wind at temperatures of about 40\,kK, in the temperature range of the He{\sc ii}-bump (the compact stellar model in Fig.\ref{structure}). Instead, the same star with a lower mass-loss rate, for example, $\rm \Mdot=5\times 10^{-4}\Msun/yr$, would find sufficient radiative acceleration to overcome gravity only at much lower temperatures, around 10\,kK, thus in the temperature range of the H/He{\sc i}-bump (the most extended stellar model in Fig.\ref{structure}). 
Fig.\ref{mdottsrel} shows, however, that when we follow a given contour in Fig.\ref{sonicdiagrammdot}, not all combinations of sonic-point temperatures and L/M of a star can sustain a stationary radiation-driven stellar wind. 

The need to transfer momentum from the radiation field to the outflowing stellar material requires another condition at the sonic point of an accelerating radiation-driven wind, that is, the opacity gradient has to be positive \citep{1999LamersCassinelli,2002Nugis}.
Figure \ref{sonicdiagrammdot} indicates that independently of the results of stellar models, a discontinuity in the sonic-point temperatures of steady-state radiation-driven stellar outflows, that is, the forbidden temperature range, at $\rm T_S\approx 20-30\,kK$. (indicated by the dashed black line in Fig.\ref{sonicdiagrammdot}). For $\rm L/M=10^{4.4}\Lsun/\Msun$, for example, no mass-loss rate implies sonic-point temperatures in this range (Fig.\ref{sonicdiagrammdot}), considering that for $\rm \Mdot>10^{-3.2}\Msun/yr$ the stellar wind is driven at $\rm T_S\gtrsim 30\,kK$, while for $\rm \Mdot<10^{-3.2}\Msun/yr$ at $\rm T_S\lesssim 20\,kK$. This is a direct consequence of the decreasing opacities for decreasing temperatures \citep[][see Fig.\ref{opacita}]{1999LamersCassinelli,2002Nugis}.   

The dashed black line in Fig.\ref{sonicdiagrammdot} therefore corresponds to the helium-minimum mass-loss rate (as in Figs.\ref{mdottsrel}\,and\,\ref{structure}) separating outflows driven to sonic velocities by the radiative force in the temperature range of the He{\sc ii}- and the H/He{\sc i}-opacity bumps.   
The fact that the increase in mass-loss rate takes place within the forbidden temperature interval while the edges of the mass-loss profile are below and above the helium-minimum mass-loss rate is the reason for the lack of stable configuration in our evolutionary models. In other words, the imposed mass-loss profile never crosses the thick black line indicating the loci of possible stationary sonic-point conditions (Fig.\ref{mdottsrel}).

The sonic-point diagram reveals that the opacity bumps reduce the L/M that is required to drive a given mass-loss rate by stellar wind in certain temperature intervals, and can also be used to predict the sonic-point temperature of a star, given the observed \Mdot and L/M \citep{2018Grassitelli}, relying solely on the atomic physics in the opacity tables. Moreover, for a given mass-loss rate, a higher L/M also implies a higher sonic-point temperature, which in the context of the S Dor variability, might explain the temperature dependence in their distribution in the HR diagram \citep{2012Vink}. 

We implicitly assumed that the Rosseland opacity is a good approximation for the flux-weighted opacity up to the sonic point \citep[which is in general the case for optically thick winds,][see also Appendix \ref{sect.opacita}]{2018Grassitelli,2019Ro}, and stress that the assumption of a purely radiation-driven dense wind is valid only for the most luminous stars near their Eddington limit \citep{1999LamersCassinelli,2002Nugis}.

\subsection{Mass-loss prescription}

 The temperature-dependent mass-loss prescription adopted for the numerical calculation in Sect.\ref{Sect.windenvelope} consists of a constant $\rm \Mdot=5\cdot 10^{-4}$\Msun/yr for $T_S>25$\,kK, followed by a continuous increase in mass-loss rate in the range $25>T_S>20$\,kK, and by a constant mass-loss rate $\rm \Mdot=1.5\cdot 10^{-3}$\Msun/yr for $T_S<20$\,kK (Fig.\ref{sonicdiagrammdot}). This is physically motivated by the mass-loss prescriptions of the Monte Carlo atmosphere models by \citet[][]{2004Smith}. Stars with an effective temperature lower than 25\,kK appear to have systematically higher mass-loss rates and lower terminal wind velocities than those above as a result of changes in the ionization equilibrium of iron atoms in particular, one of the most important elements with respect to the line-driving of stellar winds \citep[i.e., the bi-stability jump,][]{1990Pauldrach,1995Lamers,1999Vink,2001Vink,2008Puls,2011Groh}. While the mass-loss rates in the top right corner of the HR diagram are still only poorly constrained and a subject of intense research \citep{2006Crowther,2008Markova,2016Petrov,2018Vink,2020Groh}, this aspect is well beyond the scope of this paper. However, for the cycle to operate, the origin of the change in mass-loss rate is of secondary importance as long as an increase takes place within the forbidden temperature range and the mass-loss rate at temperatures below 20\,kK is above the threshold. As for the mass-loss rate adopted for $T_s>25\,$kK, the absolute value is of marginal relevance for the instability, implying that a significantly more pronounced difference in mass-loss rates associated with the bistability mechanism would not have altered the model phenomenology discussed in Sect.\ref{Sect.windenvelope}.  
 
 In our evolutionary calculations, we adopted the sonic-point temperature of the atmosphere models by \citet{2004Smith} rather than the effective temperature as the independent variable. We also uniformly increased it to clearly cross the helium-minimum mass-loss rate of the adopted stellar model. Compared with the maximum empirical mass-loss rates of AG Carinae \citep[Fig.\ref{minagcar}][]{2002Vink}, the imposed mass-loss is higher by approximately a factor 5. For clarity, we remark that the mass-loss rates in Fig.\ref{minagcar} have not been artificially increased, and that the increase is solely motivated by the relatively high specific threshold mass-loss rate of the adopted proof-of-concept stellar evolutionary model at that stage of evolution.

\section{Opacity tables}\label{sect.opacita}
\begin{figure}
\resizebox{\hsize}{!}{\includegraphics{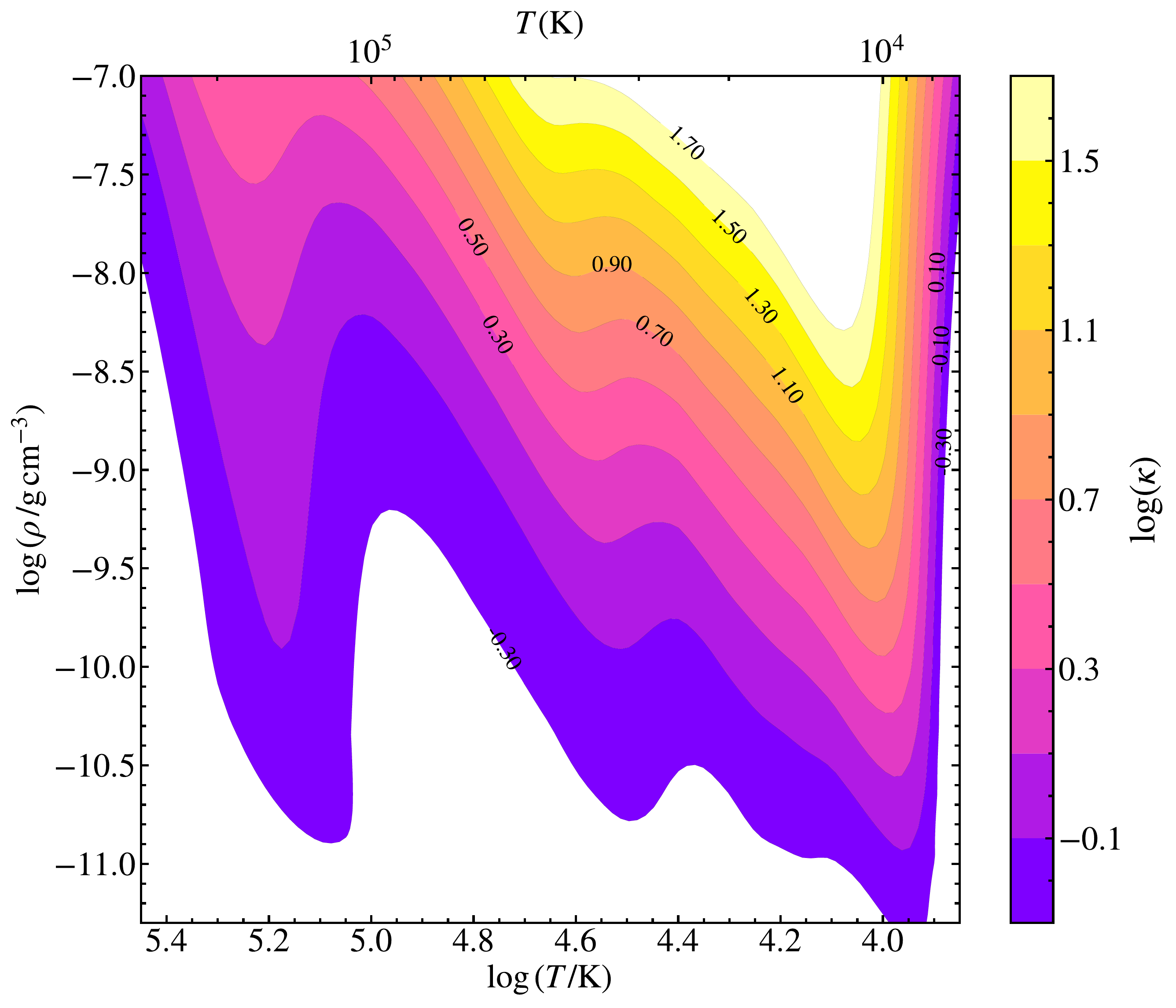}}
\caption{Opacity, log($\kappa$), from our modified OPAL tables (see the color bar on the right side) as a function of density and temperature. The opacities correspond to a chemical composition X=0.5, Y=0.48, Z=0.02, including a turbulent line-broadening of 10 \kms. The contour shapes highlight the Fe-, He{\sc ii}-, and H-opacity bumps around log($T$/K)$\,\approx 5.2,\, 4.5,$ and 4, respectively.}
\label{opacita}
\end{figure}

Our stellar models and the sonic-point diagram provide an illustrative picture of the sonic-point conditions of massive stars with optically thick winds, which has already proven to be consistent with the results from radiation-hydrodynamic stellar wind models for early-type WR stars \citep{2018Grassitelli,2020Sander}.
However, for the sonic-point diagram as well as our stellar models in Fig.\ref{tstime} to be representative of the realistic physical conditions at the sonic point of massive stars, we further assumed that the sonic point is located deep enough in the stellar atmosphere, such that matter and radiation are near local thermodynamic equilibrium (LTE). This is because the OPAL tables are derived under the assumption of LTE \citep{1996Iglesias}, and therefore they might not be representative, to a certain extent, of the stellar flux-weighted opacity in optically thin conditions. 
For the stellar models in Fig.\ref{tstime}, the Rosseland continuum optical depth of the wind ranges from $\approx 2$ in the hot minimum brightness phase to $\approx 10$ in the cool maximum brightness phase (see Fig.\ref{RsTs}). 

The OPAL tables do not account for additional effects such as line-deshadowing associated with velocity gradients, magnetic fields, or rotational broadening \citep{2008Puls,2017SimonDiaz}. 
We also modified the classical OPAL opacity tables to include the otherwise neglected effects on the line shape from the turbulent line-broadening ($\upsilon_{turb}$) with the radiative transfer code CMFGEN \citep{1998Hillier}. We were motivated by the empirical evidence of strong turbulence in the atmospheric layers of hot massive stars \citep{2017SimonDiaz,2018Jiang}, and the discrepancy between atmosphere models that include such turbulent broadening \citep[e.g.,][]{2008Grafener,2011Groh,2020Sander}, while the OPAL opacity tables do not.
For Figs.\ref{opacita}, \ref{sonicdiagrammdot}, and our numerical calculations, we adopted $\upsilon_{turb}=10\,$\kms, while for Fig.\ref{minagcar} we adopted $\upsilon_{turb}=20\,$\kms, following the atmosphere models for AG Carinae by \citet{2011Groh}. The inclusion of turbulent line-broadening results in more pronounced opacity bumps. These in turn lower the helium-minimum mass-fluxes (and mass-loss rates) by $\approx$ 0.3 and 0.6 dex for 10 and 20\,\kms, respectively, compared with the classical OPAL tables. 

The appearance of localized overdensities in the stellar winds, that is, clumping, is also expected to alter the physical conditions in the stellar atmospheres \citep{2008Puls}. If present at the base of the stellar wind of these massive stars, clumping and porosity can affect the mean opacity and consequently, the local radiative force. This has the potential of inducing further temperature-dependence in the radiative acceleration during the S Dor cycle, especially across the bistability jump \citep{2011Muijeres,2014Petrov,2019Driessen}. However, as long as a forbidden temperature range emerges, we do not expect this to fundamentally alter our physical scenario.

The smaller and irregular brightness microvariations of LBVs (cf. Fig.\ref{vismag}) likely arise as a result of such turbulent and inhomogeneous atmospheres and their complex time-dependent wind structures. Vigorous convection from both the iron and helium subsurface convective zones can lead to stochastic brightness variations on a dynamical timescale \citep[as shown by the $\approx\,0.1$mag variations on a timescale of few days in the multidimensional simulations by][]{2018Jiang}. These fluctuations, superposed on the longer S Dor cycle, could also affect the sonic-point conditions and the outer stellar layers, indirectly affecting the putative regular large brightness variations of the S Dor cycle. 

These uncertainties might in general limit the validity of the sonic-point diagram in predicting the sonic point conditions of LBVs accurately (beyond the scope of this manuscript), and lead to uncertainties in the sonic-point conditions of our stellar models. Nonetheless, we assume that none of these effects is crucial in launching the winds of LBVs because we do not expect that they fundamentally alter the depicted physical scenario. We instead expect that line-deshadowing becomes crucial to the acceleration of the stellar wind in the supersonic optically thin outer wind \citep{2008Puls}.

\section{Initial mass range and variability involving other forbidden regions}\label{sect.initmassrange}

We can extend the use of the helium-minimum mass-loss rate, together with the mass-loss prescription from detailed stellar wind models \citep{2001Vink,2004Smith}, to estimate the initial-mass range of solar metallicity main-sequence stars undergoing an S Dor cycle.
The expected L/M interval is shown in Fig.\ref{minagcar}, that is, between $\approx 4.39$ and $\approx 4.47$, indicating the L/M interval of stars that as they cross the bistability temperature redward are predicted to increase their mass-loss rate enough to exceed the helium-minimum mass-loss rate. In this we implicitly assumed that during the evolution of these stars, exceeding this threshold for the first time is sufficient to trigger the instability. It roughly and conservatively corresponds to stars with initial masses between 80 and 150\Msun, and it is restricted to stars in their main-sequence phase experiencing transitions between He{\sc ii}- and H/He{\sc i}-driven winds. This estimate, as well as the test calculations shown in Fig.\ref{tstime}, are representative of the more luminous LBVs. However, we emphasize that several physical parameters affect the exact range (e.g., rotation, mass-loss prescription, radius, and evolutionary stage), and this range does not include the wind-envelope instability involving the transition He{\sc i}$\rm \leftrightarrow$H-bump.

The He{\sc i}-bump at $\approx15$\,kK can separate from the H-bump following an increase in the turbulent line-broadening and in the He-mass fraction at the surface, for instance. This opens another forbidden temperature range around 12\,kK, between the He{\sc i} and the H-bump (see the constant opacity contours in Fig.\ref{opacita}). Combining this new forbidden temperature range with the second or cooler bistability jump \citep{2016Petrov} might well explain, mutatis mutandis, the LBV variations of the cooler and less-luminous group of LBVs \citep{2004Smith,2012Vink,2020Koumpia}, consistent with our novel wind-envelope instability. 
It is likely that as speculated by \citet{2004Smith}, this group of stars exhibit a different flavor of variability, triggered following a red or yellow supergiant phase. The idea that post-main-sequence H-rich stars evolve along the red supergiant branch increasing their L/M, until they reach the threshold helium-minimum mass-loss rate would explain the apparent metallicity-independent upper luminosity limit of red supergiants as an empirical limit for the triggering of the S Dor instability \citep{2004Smith,2018Davies}, as well as the formation of shells and nebulae due to the changing wind properties  \citep[e.g.,][]{2020Koumpia}.
For post-main-sequence stars, we expect a significantly lower initial-mass limit because the L/M is higher (higher by more than an order of magnitude than the main-sequence) reached in the late phases of stellar evolution \citep{2011Brott,2012Langer}. This suggests that, indeed, several massive stars progenitor of type II supernovae might have experienced an S Dor phase shortly before their core collapse \citep[e.g.,][]{2018Boian}.

We note that for the most massive stars at very high L/M (i.e., $\rm L/M\gtrsim 10^{4.5}\Lsun/\Msun$), the peak opacity of the Fe-bump can be higher than that of the He{\sc ii}-bump \citep{2018Grassitelli}, implying an iron-minimum mass-flux that becomes lower than the corresponding helium one. Therefore, an extended cool and very massive star experiencing an increase in mass-loss rate (either during its evolution or due to high mass-transfer rates in a binary system) might exceed the iron-minimum mass-loss rate earlier than the helium one. It would consequently develop a supersonic flow deep within the star while still being cool and extended, which might induce the eruptive loss of the \citep[potentially massive,][]{2015Sanyal} unstable overlying stellar envelope.

Of the massive stars classified as LBVs, a handful show a distinct variability from the classical S Dor, with blueward excursions toward early WR subtypes (i.e., WN3--8).
From our study, we expect that variability and phenomenology similar to that of classical S Dor LBVs could be present in other regions of the HR diagram. HD5980A \citep{2010Koenigsberger}, MCA-1B \citep{2020Smith}, and GR290 \citep{2011Polcaro,2012Clark,2020Maryeva} are possible representative examples of transitions involving the sonic points in the temperature range of the iron opacity bump, as indicated by the model atmospheres by \citet{2011Georgiev}. Their variations appear to share a quiescent phase at surface temperature $\approx30\,$kK with the classical S Dor variables, but experience blueward rather than redward excursions as they potentially cross the iron- instead of the helium-minimum mass-loss rate \citep[see][who first introduced the iron-minimum mass-loss rate]{2018Grassitelli}. In this context, outward-decreasing opacities appear between the Fe- and He{\sc ii}-bump at $T_S\approx90-160\,$kK.

Moreover, Figs.\ref{sonicdiagrammdot}\,and\,\ref{opacita} show decreasing opacities starting from $\rm T_S \lesssim 8-9\,$kK, following the peak opacity associated with the recombination of hydrogen. This forbidden range appears at temperatures comparable with a narrow region in the HR diagram that appears to be loosely populated by massive hypergiant stars, the so-called yellow void \citep{1998deJager}. Yellow hypergiant stars near this temperature range have extended envelopes close to the Eddington limit and show poorly understood variability in their surface temperatures and mass-loss rates \citep{1998deJager,2020Koumpia}. Stellar atmosphere models find a lack of radiative acceleration around this temperature \citep{1995Nieu}.
This is consistent with our expectations from Fig.\ref{sonicdiagrammdot}, as we do not expect to find massive stars near the Eddington limit with stellar winds launched within forbidden temperature ranges, serving as further observational evidence of the validity of our approach.
Our wind-envelope instability might be a viable explanation for the variability of stars such as $\rho-$Cassiopeiae, and more generally, the atmospheric instability of stars near the yellow void.

\end{document}